# Transverse Mode Instability: A Review

CESAR JAUREGUI,[1,*] CHRISTOPH STIHLER,[1,3] AND JENS LIMPERT[1,2,3]

[1]*Institute of Applied Physics, Abbe Center of Photonics, Friedrich-Schiller-Universität Jena, Albert-Einstein-Str. 15, 07745 Jena, Germany*
[2] *Helmholtz-Institute Jena, Fröbelstieg 3, 07743 Jena, Germany*
[3] *Fraunhofer Institute for Applied Optics and Precision Engineering, Albert-Einstein-Str. 7, 07745 Jena, Germany*
*\*cesar.jauregui-misas@uni-jena.de*

**Abstract:** This work presents a review on the effect of transverse mode instability in high-power fiber laser systems and the corresponding investigations led worldwide over the last decade. This manuscript includes the description of the experimental observations, the physical origin of this effect, as well as some of the proposed mitigation strategies.

## 1. Introduction

The development of the LASER in 1960 [1] started a revolution that has permanently changed our life. One of the reasons for the lasting impact of this technology is that the coherent radiation of laser sources made new applications possible and improved the performance of many others. This way, over the years, fields as widely different as medicine, industry, science and entertainment have found uses for lasers. After more than 50 years of success, the current ubiquity of lasers might misleadingly make us think that the end of this revolution is in sight but, in fact, the contrary is true: the demand for lasers keeps on increasing and they outline themselves as the corner stone of new technologies and processes yet to be developed.

Even though nowadays a world without lasers is unimaginable due to the deep economic, scientific and social impact of this technology, in the early days there were very few applications for this kind of coherent radiation. In those days lasers could only conquer the imagination of the people and they rapidly became icons of the pop culture making appearance in science fiction books and films. However, no significant practical applications were found and the development of lasers remained an academic exercise without further relevance. This led to the curious circumstance that the performance of lasers was, for a long time, ahead of the needs of the slowly developing applications for this technology. That is why the lasers were popularly labelled as "a solution looking for a problem". However, with time, the number of applications for lasers kept on increasing and the demands on the laser performance started to catch up with the laser development. In fact, at the present time it can be stated that the laser development is mainly application-driven and that there are even some important applications [2] that can only be addressed with a further development of current laser technology. This explains why the interest and commitment to this technology remains today as strong as ever.

This application-driven development has led to the scaling of three main parameters of the laser radiation: the average power, the pulse energy and the peak power (achieved mainly by pursuing a shortening of the pulse duration). In the early days, most of the developing efforts were focused on increasing the average power of the laser radiation. This development soon found the thermal limits of the original rod-laser architectures. In fact, it was observed that when increasing the output average power in the early lasers, the output beams got distorted and their spatial coherence was strongly reduced, which severely restricted the use of laser sources in real-world applications. In order to solve this problem, new laser architectures were developed, which tried to mitigate the thermal problems by changing the geometry of the active medium. These geometry changes were aimed at obtaining a larger surface-to-active-volume-ratio, which allows for a better evacuation of the heat. The most successful among these new architectures/geometries were the slab, the thin-disk and the fiber lasers [3–5]. From these, in fact, the fiber-laser architecture is the one that currently holds the record in the highest emitted average power with diffraction-limited beam quality (6 kW out of a Yb-doped silica fiber [6]). This fact is particularly remarkable taking into account that fibers are made of glass, which has a thermal conductivity that is roughly 10-times lower than that of the crystals other laser active materials are made of (e.g. slabs and thin-disks). In this context, for the fiber to hold the current record in diffraction-limited average-power laser emission is a clear statement about the outstanding properties of optical fibers as a laser active medium. The advantages of fibers arise not only from their thin, elongated geometry but also from the fact that they are the only high-power active medium that possesses a waveguide structure. Thereby, fiber laser systems are able to deliver diffraction-limited beams even under a high thermal load in a wide average power range without requiring any complex optical arrangements, unlike other competing solid-state laser technologies [3,4].

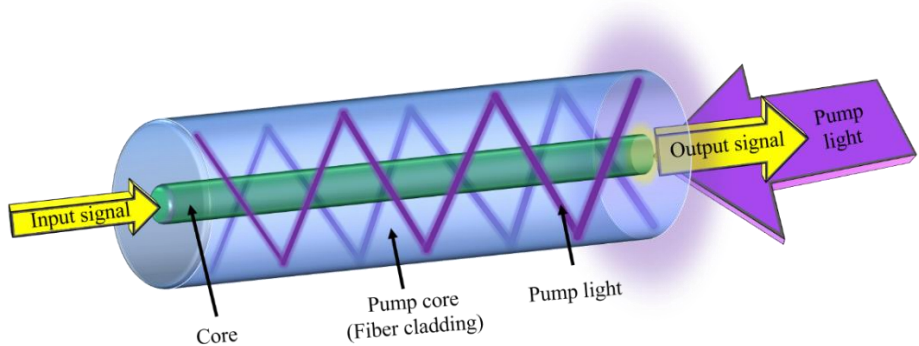

Fig. 1. Schematic representation of an active optical fiber for high-power operation. Adapted from Nat. Photonics 7, 861 (2013) [7].

An active fiber for high-power operation has, as schematically shown in Fig. 1, a signal core doped with laser active ions (such as $Yb^{3+}$, $Er^{3+}$ or $Tm^{3+}$), which is surrounded by a cladding, also known as pump core. In an amplifier configuration, the seed signal is coupled in the signal core and the pump light (characterized by a high power and low brightness) in the much larger pump core/fiber cladding. This "double-clad" concept [8] allows for the interaction between the pump photons and the laser active ions in the signal core over very long lengths. This results in a very high single-pass gain and in a distribution of the thermal load along the fiber. The latter further improves the thermal handling capabilities of active optical fibers.

Apart from their outstanding thermal properties, optical fibers offer additional practical advantages. For example, their very high single-pass gain allows for simple and robust optical setups, which can even be used outside of a protected laboratory environment. Additionally, fiber laser systems are fairly insensitive against environmental perturbations, e.g. temperature changes. Moreover, fiber laser systems are energetically very efficient and, in many cases, maintenance free. These characteristics have contributed in recent years to a rapid increase in popularity of fiber laser systems for industrial applications. However, in spite of these good prerequisites for the generation of high average powers in continuous wave (CW), the pulsed operation of these systems is challenging. On the one hand, the stored energy in an active fiber is relatively low (at least in comparison with other solid-state laser active media) and, on the other hand, the fiber geometry, with its small and long signal core, favors the onset of non-linear effects that can significantly degrade the quality and peak power of the emitted pulses.

The non-linear effects that represent the strongest limitation for the generation and amplification of high peak-power laser pulses in an active optical fiber are: Raman and Brillouin scattering, self-phase modulation (SPM) and self-focusing [9]. The latter depends on the peak power of the pulses and represents the ultimate limit for this parameter in a fiber, since values higher than 4 MW (with linearly polarized light) or 6 MW (with circular polarization) [10] lead to the destruction of the fiber material. In contrast to self-focusing, the Raman and Brillouin scattering and the SPM depend on the intensity of the light in the fiber core and can, therefore, be effectively mitigated by reducing the intensity of the signal pulses. In order to achieve this, it is possible to stretch the laser pulses in the time domain (using the so-called "Chirped Pulse Amplification" technique, CPA [11]) and/or to increase the size of the signal core in the fiber. Whereas the first approach mitigates all non-linear effects in pulsed systems, the second one can be used both in CW and in pulsed operation, even though it is only effective against intensity-dependent non-linear effects (such as SPM, Raman and Brillouin scattering). In spite of this, the second approach is particularly attractive from the practical point of view since this strategy usually results in fiber designs with a higher pump absorption. This, in turn, allows reducing the active fiber length which further mitigates non-linear effects. The disadvantage of large signal core diameters is that they are usually able to guide the light in several transverse modes, which negatively impacts the emitted beam quality and stability. In order to ensure effective single-mode operation with the highest beam quality, advanced designs have to be used in large core fibers [12–16].

For the reasons mentioned above, the active fibers for high-power operation have become progressively shorter as their signal core diameter has increased. This development has led to an unparalleled exponential evolution of the average and peak power of fiber laser systems, which has been sustained over two decades [7]. However, this impressive evolution has resulted in an extreme increase of the thermal load in these systems, which has resulted in the onset of thermal effects in active fibers. There are two main manifestations of thermal effects in fibers: the first one is the shrinking of the mode size and the second one is the so-called transverse mode instability (TMI). The first of these effects, i.e. mode shrinking, describes the progressive reduction of the mode field diameter with increasing output average power, which intensifies the impact of non-linear effects and, therefore, partially negates the benefits of a large signal core. More dramatic and wide-reaching are the consequences of TMI: the quality and stability of the beam emitted by a fiber laser system are suddenly reduced once that a certain average power has been reached. This system-specific average power limit is called TMI threshold and, depending of the fiber laser system, it usually is anywhere between 100 W to several kWs. The sudden onset of the beam fluctuations together with the dependence on the average power point towards TMI being the first representative of a thermally-induced non-linear effect observed in an optical fiber. This alone sets TMI apart from any other known effect happening in a fiber laser system. On the other hand, this also implies that the laboriously gained knowledge about the dependencies, behavior and mitigation of conventional non-linear effects cannot be applied to TMI. In other words: TMI was, at the time of its discovery, scientific new ground.

It took a short time until TMI revealed itself as the strongest limitation for the average-power scaling of fiber laser systems. This can be schematically seen in Fig. 2, where it can be observed that no increase in the record average power of nearly diffraction-limited, single-channel Yb-doped fiber laser systems has been achieved in the decade elapsed since the discovery of TMI. This explains why this topic has attracted so much attention not only from the scientific community but also from industry. Such amount of attention has led to a significant effort being put worldwide to understand and mitigate TMI, since no less than the reputation of fiber laser systems is at stake. At this point, it is only fair to mention that the hard work of many scientists around the world is starting to give fruits, and we have recently seen a rapid increase of the output average power of nearly diffraction-limited, Yb-doped fiber-laser systems with powers around 5 kW being reported [17,18]. This means that the expansion of Fig. 2 with new points can be expected soon.

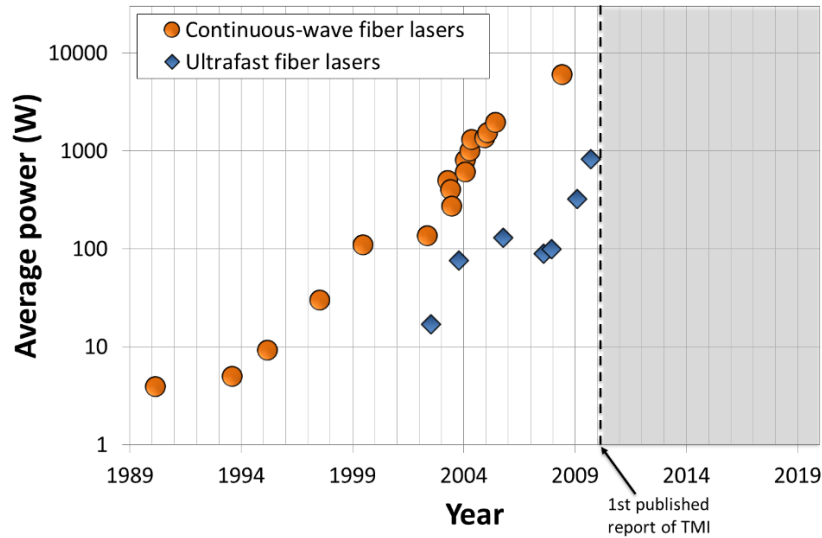

Fig. 2. Evolution of the record output average power of diffraction-limited, single-channel Yb-doped fiber laser systems (both in continuous-wave and in ultrafast regimes) over the years. Adapted from Nat. Photonics 7, 861 (2013) [7].

In the following chapters several aspects of TMI will be discussed and the most important results will be shortly presented. In particular, the first experimental observation of TMI and the temporal dynamics of this effect will be analyzed. Besides, the theoretical explanation for the physical origin of TMI will be detailed and the interplay between TMI and photodarkening [19] will be illustrated. Finally, mitigation strategies for TMI and guidelines to increase the average power of fiber laser systems will be presented and analyzed. It is important to stress that the research on TMI is still ongoing and some of the points are still hotly discussed in the community. When such topics are presented in this manuscript, we will try to point them out and provide the different alternative theories. However, it should be mentioned that, inevitably, this review is ultimately written from the point of view of the authors, which might not always coincide with the views and opinions of other scientists.

## 2. Experimental observations

The first reported observations of TMI date back to the year 2009/2010, as it was seen how the Gaussian output beam of a high-power, rod-type fiber laser system suddenly broke apart and started to fluctuate once that a certain average power level was reached [20,21] (see Visualization 1 [21]). The publication of this experimental observation marks the beginning of the research on TMI worldwide.

As already mentioned, in the first experiments it was possible to see that the formerly Gaussian beam emitted by a fiber laser system suddenly changed its intensity distribution once that a certain average output power had been reached. This change of the intensity distribution of the beam, observed with a normal CCD-camera, was not stable in time and pointed towards the appearance of higher-order modes (HOMs) at the fiber output. Such behavior is schematically illustrated in Fig. 3 for four different states (i.e. intensity distributions) of the output beam, where it can already be recognized that the distribution of the different transverse modes at the output of the fiber is not stable with time. Thus, this observation led to this effect being referred to as "transverse mode instability" (TMI).

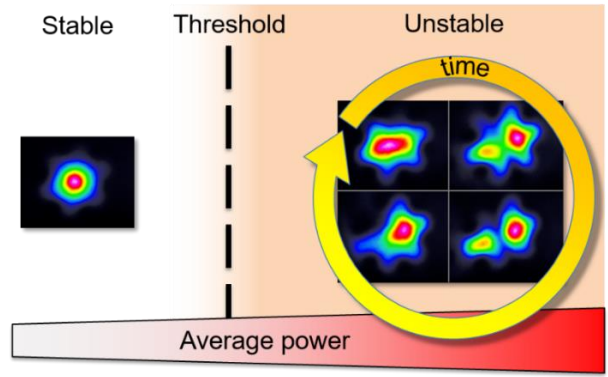

Fig. 3. Schematic representation of TMI. Adapted from Nat. Photonics 7, 861 (2013) [7].

The beam fluctuations originated by TMI are significantly faster than the frame rate of conventional cameras (i.e. 25 Hz). Therefore, in order to carry out a detailed study with the goal of learning more about the physics behind this intriguing effect, it was necessary to use a high-speed camera [22]. Therewith it was possible to obtain, for the first time, fully spatially- and temporally-resolved measurements of TMI (see Visualization 2 [22]). Three conclusions could be drawn from these results: first, the beam fluctuations happen on a ms-time scale; second, the beam is at all times the result of the coherent superposition of two or more transverse modes; third, between the stable beam operation state (i.e. below the TMI threshold) and the unstable beam operation typical of TMI (i.e. above the TMI threshold) there is a third state, characterized by periodic beam fluctuations, known as transition region. These findings provide some important insight in the physical origin of TMI. From the measurements it seemed that TMI was the result of a phase-matched process that allows for an energy transfer between two or more transverse modes and which becomes stronger with increasing average power. Additionally, the sudden onset of the beam fluctuations at a certain average-power threshold suggested that TMI arises from a non-linear effect which, as opposed to all other previously known classic non-linear effects in an optical fiber, does not depend on the peak power of the light but on its average power instead.

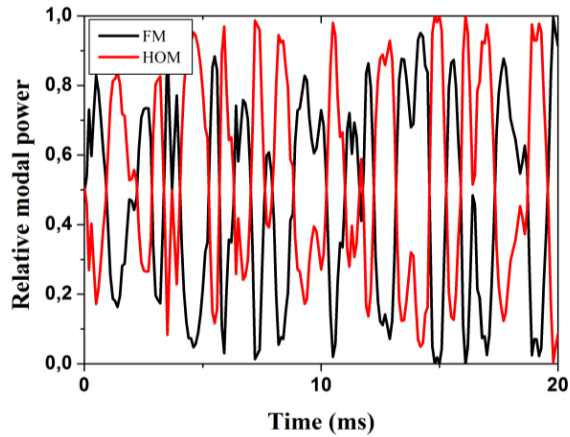

Fig. 4. Energy transfer between two transverse modes in a fiber (FM and HOM) during TMI (observed with a high-speed camera and reconstructed with a modal decomposition algorithm). Adapted from Opt. Lett. 36, 4572 (2011) [23].

Even though the most likely cause for the fluctuations of the output beam seemed to be an energy transfer between the fundamental mode (FM) and one or more HOMs in an optical fiber, a second explanation was also possible: that the fluctuations are only a result of the change of the relative phase between these transverse modes. In order to answer this central question, of wide-reaching consequences for the understanding of the physical origin of the effect, the individual frames of the high-speed video were analyzed and the mode content in them reconstructed. This was achieved by using an intensity-based mode reconstruction algorithm on the intensity pictures captured in the video frames [23]. Using this procedure it was possible to retrieve the temporal evolution of the relative modal energy distribution and phase required to recreate the changes of the output beam. The result of this analysis, presented in Fig. 4, shows clearly that there is indeed an energy transfer between the different transverse modes of the fiber after the onset of TMI. As can be seen in Fig. 4, an almost complete energy transfer between the FM and a HOM is possible on a millisecond time-scale. Besides, the relative phase between the modes also changes in a similar fashion. Therefore, it can be concluded that the apparently chaotic fluctuations of the output beam are a result of the combination of this modal energy transfer and the change in the relative modal phase. Please note that the term "chaotic" here is loosely employed to describe the non-predictable nature of the fluctuations without implying that the underlying non-linear dynamics of the process are chaotic in the mathematical sense.

The videos captured with a high-speed camera contain the full spatial and temporal information of the beam fluctuations but their length is restricted to just a few seconds, since the acquisition time of the camera is limited by its inner storage capacity. This fact makes long-time stability studies difficult. Besides, those cameras are quite expensive. Both of these problems can be solved in an elegant way by replacing the high-speed camera by a photodiode (PtD). For this approach to work, though, it is imperative that the aperture of the photodiode is smaller than the size of the imaged output beam [22]. The main drawback of this technique is that it gives up the detailed information of the modal composition of the beam but, in exchange, the measurement setup becomes significantly simpler and cheaper and long-time acquisitions are made possible. Even though when using the PtD the modal information is lost, the decisive detail is that the complete frequency information of the beam fluctuations remains intact and can be analyzed by applying a Fourier transformation to the temporal PtD trace. In short, this measurement technique represents an easy and elegant way to characterize TMI, which has become now standard.

With the help of the measurement based on the PtD it was possible to obtain a detailed characterization of the three average-power dependent operation states of a fiber laser system: stable, transition and chaotic. The PtD traces corresponding to any of these states are very different from those of the other states, as can be seen in Fig. 5. In the stable state (i.e. below the TMI threshold) the PtD trace shows a horizontal line, which is the trademark of a static, stable beam. In the transition state (for average powers slightly above the TMI threshold), the PtD captures a sinusoidal trace, which indicates a periodic fluctuation or change of the beam. Finally, in the chaotic state (for an average power significantly above the TMI threshold) the PtD trace shows random/chaotic fluctuations, which imply complex changes of the beam profile.

The standard deviation of the PtD traces is used to quantitatively describe the beam fluctuations with a single number. Thus, it is possible to plot the standard deviation of the PtD traces against the average output power of the system to obtain a graph similar to that shown in Fig. 5, which characterizes the evolution of the stability of the beam of a fiber laser system.

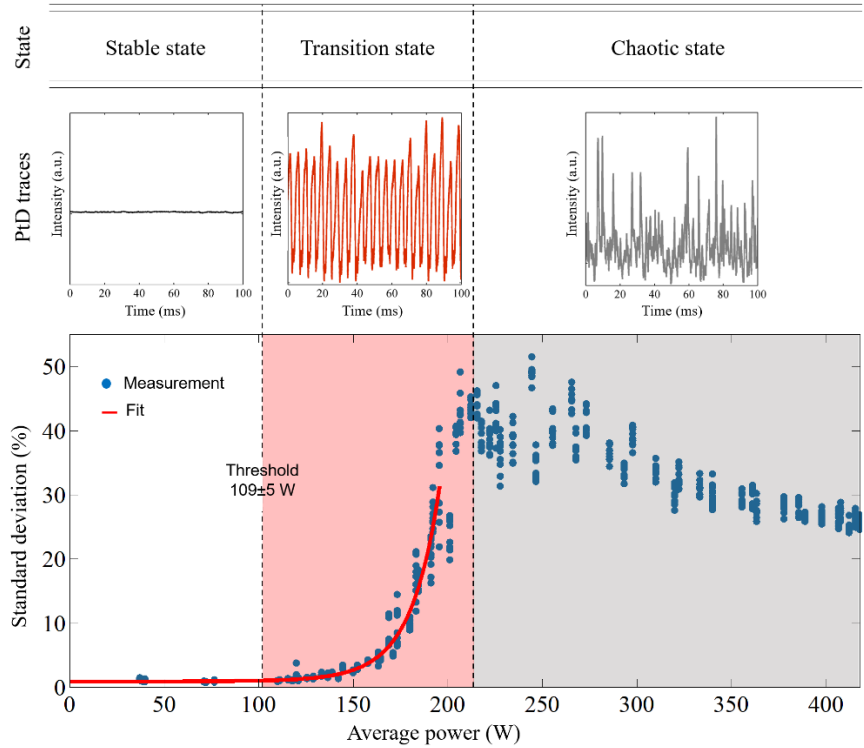

Fig. 5. Evolution of the stability of the beam emitted by a fiber laser system as a function of the average output power. The graph shows the evolution of the standard deviation of the PtD traces with increasing average power (bottom plot). Additionally, the three operation states of a fiber laser system are highlighted and their characteristic PtD-traces are shown (upper plot).

As can be seen in Fig. 5, the standard deviation shows a nearly exponential evolution between the stable and the transition states, which can be fitted with a function (red solid line in Fig. 5) to obtain a quantitative definition of the TMI threshold [22]. Thus, it was proposed that the TMI threshold is reached when the local derivative of the fitted evolution of the standard deviation with the average power becomes 0.1 ‰/W. For example, according to this definition, in Fig. 5 the TMI threshold would be 109 W. This definition of the threshold guarantees that the beam profile remains stable at least until that output average power. Furthermore, a quantitative definition of the TMI threshold such as the one proposed above ensures the comparability of the thresholds of different fiber laser systems. Please note that this definition of the threshold, even though widely used, is not the only one. For example, there are some systems, especially those using smaller and/or longer fibers, in which the transition between the stable and the chaotic states is fairly abrupt and, therefore, it is difficult to fit an exponential function there [24]. Thus, some authors have opted for simply defining the threshold as the power in which this abrupt, step-like change in the stability of the beam happens [25]. Other authors prefer to use a definition of the TMI threshold based on the $M^2$ (e.g. [26,27]), even though this definition can be less reliable than the previous ones since the degradation of the $M^2$ can be static, i.e. unrelated to TMI.

The behavior of the beam at output average powers significantly above the TMI threshold has also been studied [28]. In this study it could be seen that, with an increasing average output power, more and more modes become involved in the beam fluctuations. As a consequence, in average, the beam approaches a super-Gaussian (i.e. "flat-top") profile at high-enough output powers.

Additionally, the dependence of the characteristic frequency of the beam fluctuations as a function of the fiber core diameter was analyzed [22]. The main difficulty of such a study is that, in order to warrant the comparability of the results, it must be ensured that fibers with different core sizes guide similar modes, which is something that usually does not happen. However, this prerequisite could be fulfilled in these experiments by employing the so-called "large-pitch" fibers (LPF) [29]. This particular fiber design has the unique property that its guiding properties are theoretically nearly independent from the fiber core size. Thus, the TMI measurements with three LPFs of different core sizes have shown that the larger the fiber core the slower the beam fluctuations are. A straightforward analysis showed that the dominant frequency of the fluctuations (in the 100's of Hz to kHz range) in each fiber was compatible with the expected thermal diffusion time in the core (i.e. the larger the core, the slower the beam fluctuations), which strongly suggests a thermal origin for TMI.

A more detailed analysis of the evolution of the oscillation frequencies of the beam fluctuations caused by TMI was done in [25]. In this work, spectrograms of the beam fluctuations were recorded, which showed that the beam fluctuations become faster at higher average powers. Moreover, an increasing number of harmonics of the main beam oscillation frequency take part in the fluctuations at higher average powers.

The suspicion that TMI could be an effect of thermal origin was reinforced in an experiment in which, by pulsing the pump power, it was shown that there is a typical build-up time of some ms for the beam fluctuations [30]. Additionally, it was shown that if the pump pulses were spaced within some tens of ms the build-up time became shorter. This points towards a long decay time for the transient changes taking place in the fiber and leading to TMI.

Additionally, even most of the work on TMI has been done in Yb-doped fibers, this effect has been predicted in Tm-doped fiber amplifiers [31] and in Raman fiber amplifiers [32]. Very recently TMI has been finally observed in Tm-doped fiber amplifiers [33] demonstrating that these systems are significantly more resilient to TMI than Yb-doped fiber amplifiers. The reasons for this robustness against TMI are not fully understood yet, even though some authors believe that cross-relaxations may play a significant role in weakening the thermally-induced grating [31]. This result, however, underlines the fact that TMI is a fundamental thermo-optical effect, which is not tied to any particular laser-active ion. In other words, regardless of the fiber amplifier technology, once that a certain threshold heat load has been reached the onset of TMI can be expected.

## 3. Theoretical description of the effect

As already mentioned above, the experimental observations suggest that TMI is the result of a thermally-induced non-linear effect, which leads to the energy transfer between different orthogonal transverse modes in an optical fiber. However, in principle, such an energy transfer between orthogonal modes can only be the result of a phase-matched process. There are not many mechanisms that allow for such a phase-matched interaction in a fiber and, at the beginning of the research of TMI, none was known with a thermal origin. However, an optical element that does allow for the exchange of energy between different transverse modes in an optical fiber is a long period grating (LPG). A LPG is a periodic or quasi-periodic modification of the refractive index of the fiber with a period that is significantly longer than the signal wavelength. In order to enable energy transfer between two orthogonal modes, the period and symmetry of the index change in the LPG have to be similar to those found in the modal interference pattern created by the beating of those two modes.

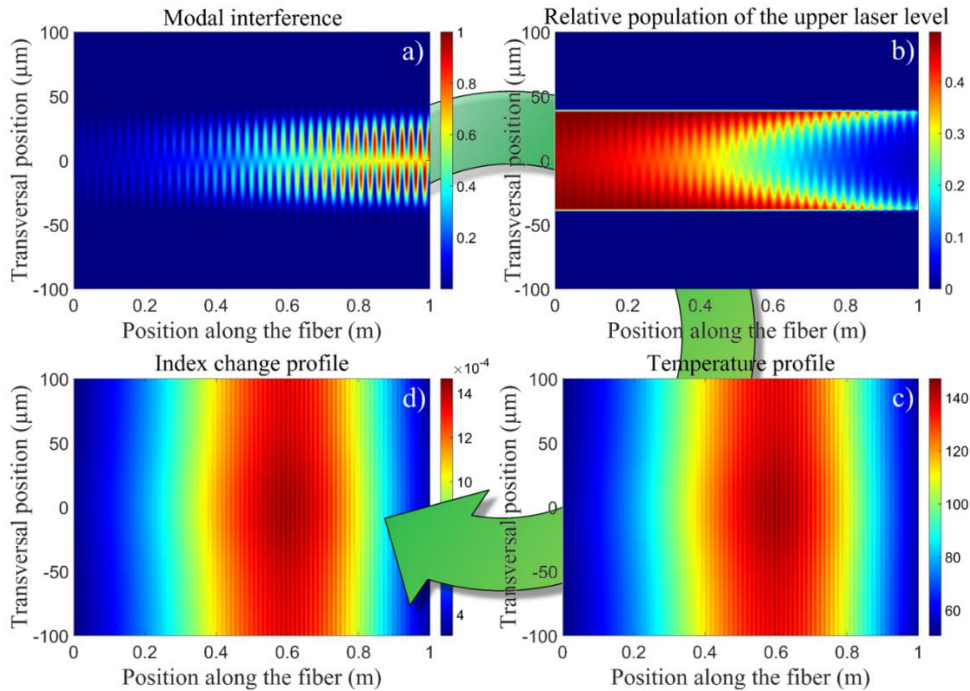

Fig. 6. Generation of a thermally-induced index grating in an active optical fiber: the periodic modal interference pattern (a) gets imprinted in the inversion profile (b). This, in turn, gives rise to a quasi-periodic temperature profile (c) that, through the thermo-optic effect, is translated into a refractive index change, i.e. a grating (d). Adapted from Nat. Photonics 7, 861 (2013) [7].

Shortly after the publication of the first observations of TMI [20,21], it was proposed that either an inversion-induced or a thermally-induced LPG could be responsible for TMI [34,35], with temperature being thought as the most likely cause. Nowadays this theory is widely accepted by the scientific community as the origin of the effect. In fact, this understanding has led to the widely established view that TMI is a manifestation of Stimulated Thermal Rayleigh Scattering (STRS) in optical fibers [36,37].

The generation of such a thermally-induced LPG can be schematically explained by using the 4-step model illustrated in Fig. 6 [7]. The different plots in this figure correspond to a simulation of a 1 m long fiber amplifier with 80 µm core diameter (from which only a diameter of 64 µm is doped with $Yb^{3+}$) and a V-parameter of 7. This fiber is pumped at 976 nm in the co-propagating direction and amplifies a signal at 1030 nm. As already mentioned before, such large-core fibers are generally multimode (i.e. support the propagation of more than one transverse mode) due to current technological limitations in the fabrication process (which impose a lower limit to the achievable V-parameter of a fiber design). This implies that, when light is coupled in the core of these fibers, usually several modes will be excited (normally these are the FM and the first-order HOM, which in a weakly guiding fiber is designated as $LP_{11}$). For example in the simulation depicted in Fig. 6 it has been assumed that the FM carries 70 % of the input energy and the $LP_{11}$ 30 % (please note that these values have only been chosen for a clear illustration of the different steps of the model, but they are not representative of a real excitation. Usually, in experiments the $LP_{11}$ only carries a few percent of the input energy). After excitation, both modes will propagate along the fiber with different phase velocities, thus creating a modal interference intensity pattern (MIP) in the process. Such an MIP shows regions of high intensity and regions of low intensity alternating with a well-defined period, which is given by the beat length of the interfering modes (Fig. 6a). Taking into account that in an active fiber the inversion is more strongly depleted in regions with higher signal intensity, it can be easily shown that the MIP will get imprinted in the inversion profile, as can be seen in Fig. 6b. This implies that the MIP gives rise to an inversion profile that is transversally inhomogeneous and quasi-periodic (in the longitudinal direction). Such an inversion profile indicates a transversal inhomogeneity in the power extraction in the fiber core which, in turn, leads to a transversally inhomogeneous and quasi-periodic temperature profile (see Fig. 6c) as a result of the heat-load created by, e.g., the quantum defect (QD). Finally, this temperature profile is translated into a refractive-index change by the thermo-optic effect (Fig. 6c), thus giving rise to a thermally-induced LPG. Moreover, since this grating has been ultimately induced by the modal interference pattern, it always has the right period and symmetry to potentially transfer energy between the interfering modes [38].

Shortly after the publication of this theory about the origin of the thermally-induced refractive-index grating (RIG), it was pointed out that a second condition needs to be fulfilled to actually enable energy transfer between the transverse modes: a phase shift between the MIP and the RIG [35]. This condition implies that the intensity maxima/minima of the MIP must be shifted longitudinally with respect to the refractive index maxima/minima of the RIG. This, in turn, leads to a movement of the RIG, because it constantly tries to adapt itself to the (shifted) heat-load generated by the MIP. Such an adaptation process is nothing but a direct consequence of the RIG being ultimately generated by the MIP. Therefore, if, for some reason, a phase shift between the MIP and the RIG appears, the latter will evolve and adapt itself to the new MIP (or, better said, to the new heat-load profile generated by it), albeit with a certain temporal delay. This delay is caused because the temperature in a fiber, and with it the thermally-induced RIG, needs a certain time to change, which is related to the thermalisation time of the fiber core. Another consequence of the movement of the RIG is that a frequency difference between the modes involved in the energy exchange appears due to the Doppler effect. This frequency difference corresponds to the main frequency of the beam fluctuations.

Thus, in a nutshell, in order to obtain the beam fluctuations caused by TMI there are two compulsory requirements that have to be fulfilled: the appearance of a refractive-index grating (thermally) induced by the modal interference pattern and a phase shift between these two entities (the RIG and the MIP), as schematically illustrated in Fig. 7. At this point it is important to note that the TMI threshold is determined by the combined strength of the RIG and the phase shift. In other words, the TMI threshold can be reached with a strong RIG and a weak phase shift or vice-versa.

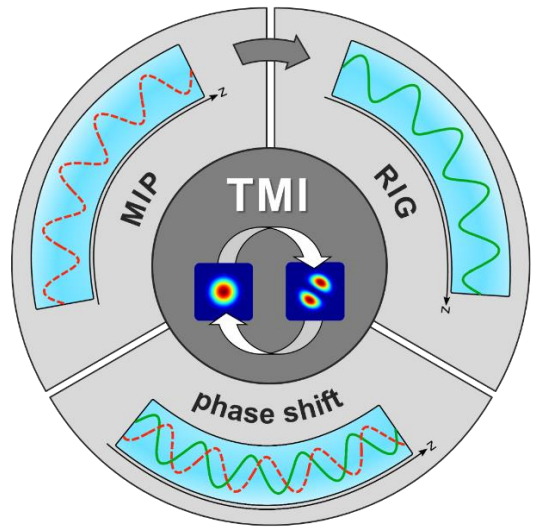

Fig. 7. There are two requirements to obtain TMI: first, the presence of a modal interference pattern (MIP) that induces refractive index grating in the fiber (RIG); second, a phase shift between these elements along the fiber.

After the discussion about the need for a phase shift between the MIP and the RIG, the next big question that has to be answered is that of its origin. This point is still under discussion in the scientific community and has not been completely clarified yet. One popular theoretical explanation proposes that the transverse modes that give rise to the MIP have slightly different central frequencies (or, in other words, the distribution of their spectral energy density is slightly different) [35] and, therefore, generate a travelling interference wave in the fiber. This moving interference pattern will create the phase shift required for the energy transfer, due to the finite speed at which the temperature profile can change and follow it. Hereby, the onset of TMI will occur when the thermally-induced refractive index changes (RIG) reach a critical strength (since in this hypothesis the phase shift is fixed, usually to the optimum value of $\pi/2$). Even though the physical plausibility of this hypothesis has been demonstrated by theoretical models [35,39,40], the physical origin of the required spectral difference between different transverse modes has not been fully explained yet. There are suggestions pointing towards quantum noise, spontaneous emission, amplitude noise, or spontaneous Rayleigh scattering [41] as possible causes for such a frequency difference.

A second hypothesis, endorsed by the authors of this review, does not require this initial spectral difference between the interfering modes. In fact, in this theoretical explanation both modes have the same distribution of their spectral energy density. Therefore, this theory states that below the TMI threshold the RIG and the MIP do not (significantly) move along the fiber (they can only jiggle around an equilibrium position due to noise-induced thermal changes in the fiber). Thus, below the TMI threshold, there is no significant energy transfer between the modes since both the MIP and the RIG remain largely in phase (because since the noise-induced movements of the MIP and RIG are small, their phase-shift will not be large) and the RIG is comparatively weak [42]. As the average power increases, however, the thermally-induced index changes (RIG) become stronger and stronger, which implies that even small phase shifts are sufficient to trigger modal energy coupling. The reason for this is that substantial modal energy transfer can be induced either by a weak RIG when it has a large phase-shift with the MIP or by a strong RIG even with small phase shifts. In other words, the sensitivity of the modal energy coupling to phase shifts grows with the average power [43]. Therefore, in the context of this theoretical explanation, a transitory change of the system, e.g. caused by a fluctuation of the pump or the signal power, can suffice to generate a phase shift between the MIP and the RIG (see section 5), which is strong enough to lead to the energy transfer between modes. The likelihood of this happening increases exponentially with the average power and, thus, at some point, the intrinsic noise of the system suffices to trigger strong modal energy transfer, i.e. beam fluctuations. As it will be shown later, recent experimental evidence [44] seems to supports this theory.

The feasibility of both theories has been already proven with the help of various numerical simulation tools [27,39,40,42,45–47]. The numerical methods that use the first theory of the origin of the phase shift (i.e. the two-frequency approach) are based on the so-called "steady-periodic" model [39,48,49]. Hereby it is assumed that the two interfering transverse modes have slightly different frequencies from the beginning, which significantly simplifies the calculation of the energy transfer and even allows for the derivation of semi-analytic formulas [36,40,50,51]. The main limitation of these models is that the predicted behavior of the beam does not fully match the experimental observations. For example, these models predict that for powers slightly above the TMI threshold there should be a stable energy transfer from the FM into the HOM. This is something that has not been observed experimentally in a free running system yet. The models that support the second theory of the origin of the phase shift are significantly more complex and involved from the numeric point of view since they are based on a three dimensional beam propagation method (BPM) coupled with the thermal-diffusion and laser rate equations [27,42,45,47]. Recently a new, simplified model belonging to this family has been presented, which can run in a personal computer [52]. These models, which do not assume any frequency difference between the transverse modes at the beginning of the fiber, are able to faithfully reproduce the existence of the three experimentally-observed operation states of a fiber laser system (stable, transition and chaotic) [27]. Therefore, it can be stated that these models are able to encompass more details of the real physics of TMI than the ones based on the first theory. In any case, regardless of their strengths and weaknesses, all these models have contributed to the rapid development of the theoretical understanding of TMI in active fiber laser systems.

Very recently a completely new approach for simulating TMI has been presented [53]. This is a high level analysis of the stability of a high-power system. What is meant by high level is that this approach abstracts itself from the details of the physical origin of TMI and simply shows that fiber laser systems become more and more sensitive to perturbations at high average powers, much in the vein of the second theory of the origin of TMI discussed above. Thus, in this analysis TMI is just the manifestation of the fiber system becoming unstable. Even though the high level of abstraction might be a problem to obtain a detailed knowledge of the physical processes underlying TMI (and, therefore, it might be only partially helpful to develop mitigation strategies), this approach allows bringing different effects into the same frame, i.e. it possesses a significantly broader view on a fiber lasers system than other models. Thus, using this model considering several limitations of a fiber laser systems, there has been an estimation of the absolute maximum power than can be extracted from a single fiber, which amounts to ~35 kW in the case of diode pumping and ~80 kW for tandem pumping [54].

Interesting as these results are, maybe the most important contribution of this new approach of theoretically analyzing TMI is the fact that it is able to explain, within the frame of a single theory, the onset of different kinds of observations. For example, there have been reports of some beam instabilities arising in high-gain, small core fibers at very low powers (several Watts) [55,56]. The authors of that work showed that such behavior could be explained with a refractive-index grating induced by a non-homogenous inversion profile arising in the fiber [57]. They also showed that such grating competes with the thermally-induced grating responsible for the high-power instabilities. This does not conflict with the common observations of TMI, since in the fibers used as main high-power amplifiers the thermal grating is usually the dominant one once the fiber is saturated. The model presented in [53] allows showing that both the inversion and the thermally-induced effects can lead to beam instabilities. Moreover, very recently the authors of that model have found some evidence of inversion and thermal driven instabilities [58] in a fiber oscillator.

## 4. TMI Threshold: Influencing parameters

This section describes some of the main parameters that influence the TMI threshold in a fiber laser system. Due to the thermal origin of TMI, possibly the most important of these parameters is the heat-load in the fiber. There are several sources of heat in an active fiber. The most important one is the quantum defect which is inherent to the laser operation, since the pump photons have (usually) a higher energy than the signal photons. This kind of heat will be labelled as "productive" heat since it leads to the generation of signal photons. There are, however, other "non-productive" heat sources in which the heat is the result of the annihilation of photons. One representative of this second kind of heat sources is the linear ground absorption, which is, nowadays, in most situations, extremely low and, therefore, its contribution to the total heat-load is usually negligible. Another "non-productive" heat source in Yb-doped active fibers is photodarkening (PD) [59], which refers to a photo-induced degradation of the active fiber. In particular, PD is an effect that results in a progressive loss of transparency of an Yb-doped fiber with the laser operation time. Even though there are still ongoing discussions about the exact origin of PD, it is widely accepted that the increasing optical absorption is a result of the appearance of color centers in the fiber. However, the exact mechanism leading to the creation of these color centers is not fully understood yet. Nevertheless, it is believed that the generation of $Yb^{+2}$ ions [60] and the emission of UV radiation as a result of multi-photon absorption processes in spurious $Tm^{+3}$ ions [61] play an important role. Regardless of their origin, these color centers are responsible for the absorption of a portion of the light (both from the pump and from the signal) that propagates along the fiber, thus creating an excess PD-induced loss. It has been empirically shown that the maximum PD-induced absorption of light at 633 nm depends nearly quadratically on the $Yb^{3+}$-ion concentration [62], but the actual PD-loss (from 0 to the maximum value given by the already mentioned quadratic dependence) depends linearly on the relative percentage of excited ions in the fiber [63]. Crucially, the energy absorbed in the color centers is transformed into heat which, therefore, leads to an increase of the total thermal load in the fiber. This implies that, in an active fiber, the PD-losses will give rise to an extra heat-load that mimics the MIP in the fiber (see Fig. 6). This PD-induced heat-load term creates a new thermally-induced index grating that is in phase with the one generated by the QD and, therefore, coherently overlaps with it. This results, in general, in an increase of the strength of the overall thermally-induced RIG and in a reduction of the TMI threshold [13,25,64–66].

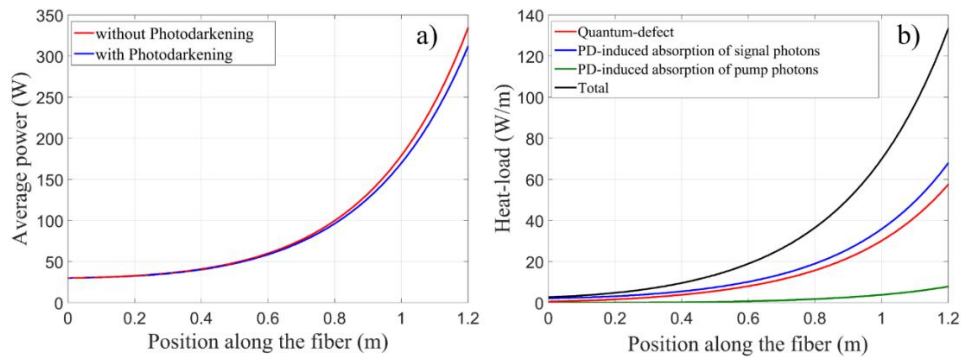

Fig. 8. Simulation of a 1.2 m long LPF-amplifier with and without PD-losses. a) evolution of the signal average power along the amplifier fiber with (blue line) and without (red line) PD. b) evolution of the thermal load along the fiber for different heat sources: quantum-defect (red line), PD-induced absorption of signal photons (blue line), PD-induced absorption of pump photons (green line) and the sum of all these contributions (black line). Adapted from Opt. Express 23, 15265 (2015) [65].

Thanks to the optimization of the composition of the active materials used in optical fibers carried out over the last decade, the PD-losses are nowadays relatively low in high-power operation with values usually well below ~2 dB/m at a wavelength around 1000 nm. This fact led, over a long period of time, to the premature but widely spread conclusion that the PD-induced heat-load in an active fiber could not be high. In order to further illustrate the reason for this assumption a simulation of a 1.2 m long fiber amplifier is presented in Fig. 8. The fiber has a 80 µm core diameter from which 64 µm are doped with $3.5·10^{25}$ $Yb^{3+}$-ions/$m^3$, 177 µm pump core diameter, 35 W of input signal at 1030 nm and a pump wavelength of 976 nm. The simulation of the evolution of the signal power along the amplifier fiber has been calculated with (blue line in Fig. 8a) and without (red line in Fig. 8a) taking the PD-losses into account (which average value in this simulation is ~1 dB/m). As can be observed in Fig. 8a, the presence of PD-losses leads to a small reduction of the average power of ~7 % at the output of the fiber. Again, this small loss of power seems to support the assumption that the PD-induced heat-load cannot be very high. However, the detailed calculations of the evolution of the heat-load along the fiber amplifier presented in Fig. 8b reveal a completely different and more dramatic picture. It can be seen that the PD-induced heat-load caused by the absorption of signal photons (blue line in Fig. 8b) is even slightly higher than the QD-induced heat-load (red line in Fig. 8b). Additionally, PD also results in the absorption of pump photons, which leads to an additional PD-induced heat-load term (green line in Fig. 8b). All in all the presence of PD-losses more than doubles the total heat-load in the active fiber, even though they only lead to an average power loss of 7 %.

The reason for this strong thermal impact of PD has to do with the fact that while QD only transforms 5-10 % of the pump-photon energy into heat in an Yb-doped fiber, in contrast, PD converts nearly 100 % of the absorbed photon energy into heat. Therefore, PD is potentially a much stronger heat source than QD in Yb-doped fiber laser systems. Thus, even low PD-losses can have a significant impact on the total heat-load of the system. This result, that has been published in [65], was very important because it revealed that it is necessary to change the way the problem of PD is approached: The optimization of PD in active fibers should not concentrate exclusively in the loss of average power any more, but it should focus on a minimization of the heat-load in the fiber.

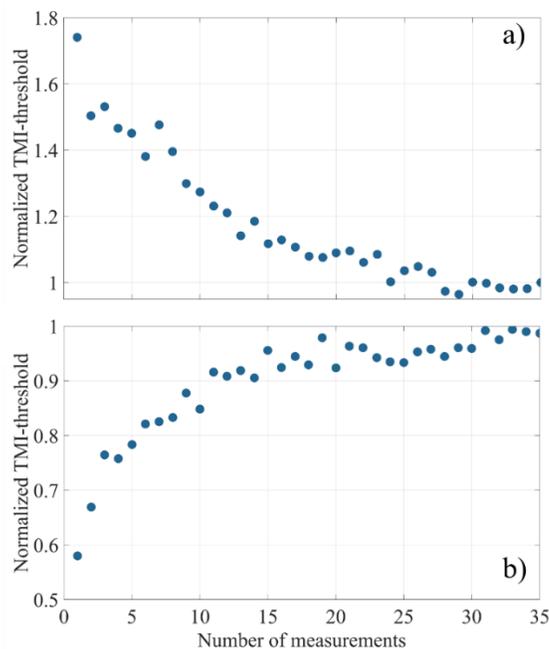

Fig. 9. Evolution of the TMI threshold with the operation time. a) for a new fiber and b) for a fiber that has been previously fully photo-degraded using a pump radiation at 915 nm. Adapted from Opt. Express 23, 15265 (2015) [65].

After the discussion presented above, it can be safely assumed that PD should impact the TMI threshold in an active fiber. The only question now is how PD will manifest itself in the context of TMI, which can only be answered by describing further experimental observations on the behavior of the TMI threshold. The first of these observations, which can be seen in Fig. 9a, is that the TMI threshold decreases with the laser operation time until an equilibrium level is reached [65]. The speed with which the TMI threshold is reduced as well as the exact power of the final equilibrium level depend on the particular characteristics of each fiber laser system. Such a behavior of the threshold remained unexplained for a long time, even though it was speculated that it could be related to PD [13,67]. The confirmation of this came after detailed experimental and theoretical evidence was presented in [65,68]. In fact, these works show that the threshold does not necessarily have to sink with the operation time, which is the usual observation, but it is also possible that, under certain circumstances, it even increases. This atypical behavior, illustrated in Fig. 9b, was measured when tracking the evolution of the TMI threshold in a fiber that has been previously fully photo-degraded. This photo-degradation, which corresponds to the maximum level of PD that can be reached in the active fiber, was achieved by operating the fiber with the highest possible level of inversion over a long time. In order to do this, the Yb-doped LPF was pumped during 1 hour with 915 nm without seed signal, which ensures a relative population of the excited state of ~80 % along the whole fiber. At the end of this process the fiber is fully photo-degraded, i.e. the PD-losses have reached their maximum.

It is known that the PD-induced losses tend towards an equilibrium state that depends on the operation parameters [63]. This equilibrium state denotes a balance between the rates of creation and destruction of color centers. This means that the PD-losses will progressively build up in a new fiber until the equilibrium state is reached. On the other hand, in a fully photodegraded fiber there are more color centers than those corresponding to the equilibrium state, so the rate of destruction of color centers will exceed that of their creation. This, in turn, leads to a progressive reduction of the PD-losses that asymptotically approaches the equilibrium state. Consequently, the TMI threshold in a fully photo-degraded fiber should increase with the operation time, if PD is responsible for the temporal changes of the threshold. Moreover, once that the new equilibrium state has been reached, the final TMI threshold power should be similar to that reached in a new fiber after its degradation (if both operate under the same conditions), i.e. that shown in the right hand side of Fig. 9a. This expected behavior describes perfectly the temporal evolution of the TMI threshold measured in the photo-degraded fiber, as can be seen in Fig. 9b, which strongly suggests that PD is the cause of these changes.

There are further observations that strengthen the thesis that PD is indeed the origin of the temporal degradation of the TMI threshold. One of them is that, after thermally annealing the TMI-degraded fibers, the TMI threshold recovered its original value before it started degrading again [66]. Likewise, in a different experiments it was reported that it was possible to revert the degradation of the TMI threshold by injecting blue light over 20 hours in the fiber [13]. The explanation for these observations is the fact that the color centers giving rise to the PD losses can be thermally and optically bleached [69].

In order to be able to interpret the experimental results better, it is useful to develop a theoretical model of the operation of a fiber laser system. The challenge in this particular case, though, was that it should include the impact of PD but, up to that point, no way to predict the expected PD-losses at the signal wavelength and in laser operation was known. In order to overcome this obstacle, the work in [68] proposed to combine the values of the PD-losses measured in [62] and [63] to obtain a heuristic formula that calculates the expected PD-losses in equilibrium state at 633 nm in alumino-silicate Yb-doped fibers. Even though this can be useful in certain circumstances, in order to model the impact of PD on the performance of a fiber laser system the PD-losses at the signal wavelength are required. Thus, a way to translate the PD-losses from 633 nm to the 1 µm wavelength region is needed. This can be done employing the results presented in [70,71], that show that the PD-losses at 1 µm are directly related to those at 633 nm by a certain constant factor. To obtain an accurate estimate for this factor, the evolution of the mode field diameter at the output of a LPF amplifier with increasing average power was simulated and it was compared with the experimental measurement presented in [72]. By considering the extra heat-load caused by PD, it was possible to retrieve a factor of 24.5 as the one giving the best fit between simulation and measurement (this factor is, however, most likely, material dependent). The introduction of this factor in the formula, which was used to predict the PD-losses at 633 nm, allows getting an estimate of this parameter at the signal wavelength. The use of this formula and its incorporation in a numerical model for Yb-doped fibers has allowed obtaining one of the first, or maybe even the first, numeric computation of the thermal impact of PD in an Yb-doped fiber during laser operation [68].

Apart from the experiments that have already been presented in this document, there are further experimental observations that highlight the important role that PD plays in determining the TMI threshold of Yb-doped fiber laser systems [68]. One of these observations is the wavelength dependence of the TMI threshold shown in Fig. 10a. This experiment was done using a 1.2 m long LPF and a seed signal power of 30 W at different wavelengths. As can be seen, the TMI threshold reaches its maximum value in this system for a signal wavelength of ~1030 nm. This is interesting because, should the QD be the only heat source in the fiber, then the TMI threshold should, in principle, increase for shorter signal wavelengths. This expectation, however, contradicts the measurement presented in Fig. 10a. In [73] it is shown, that at shorter wavelengths a lower level of gain saturation could modify this behavior, but it should not lead to the strong decrease in the TMI threshold between 1010 nm and 1030 nm observed in Fig. 10a). Hence, the experimental observation implies that there must be an additional wavelength-dependent heat source in the fiber. Thus, in order to determine the nature of this heat source this experiment was simulated with the model described in [68] obtaining a good agreement with the measurements, as can be seen in Fig. 10a. These simulations have revealed that PD is, in fact, the additional heat source that leads to the unexpected wavelength dependence of the TMI threshold. Moreover, the simulation results indicated that the absolute contribution of PD to the total heat-load of the fiber is comparable to that of QD. According to that model, the reason why the maximum value of the TMI threshold is reached around 1030 nm is related to the fact that at this wavelength the inversion level in the fiber is at its lowest, which leads to the lowest PD losses. For shorter wavelengths the QD-induced heat-load sinks but the transparency inversion (i.e. the inversion level required for a gain of 1) is higher. This leads to higher PD-losses and to a higher PD-induced heat-load that overcompensates the lower QD-induced heat-load and results in a lower TMI threshold. For wavelengths longer than 1030 nm both the QD and the PD-losses increase. The reason for the increase of the PD losses is that the emission cross-section decreases at longer wavelengths and, therefore, a higher inversion is required to achieve the same gain. This explains the reduction of the threshold in that wavelength region.

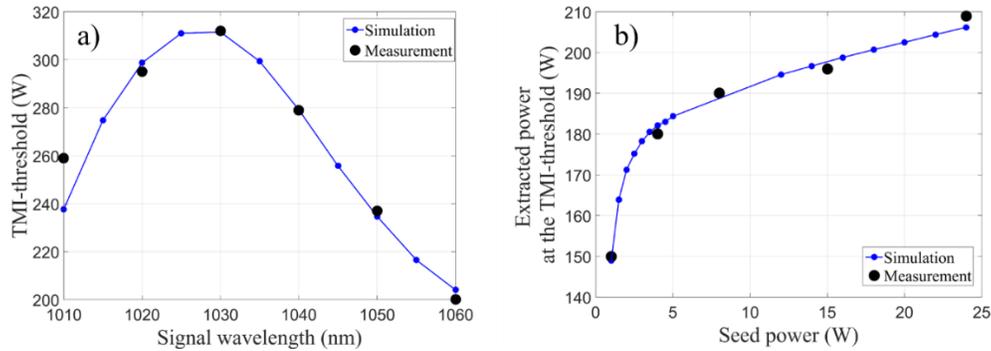

Fig. 10. Dependence of the TMI threshold with a) the signal wavelength and b) the seed average power. The black points correspond to the experimental measurements and the blue solid line shows the prediction of the model. Adapted from Opt. Express 23, 20203 (2015) [68].

The second observation that revealed the strong influence of PD on TMI is related to the non-linear dependence of the TMI threshold on the signal seed power shown in Fig. 10b. This experiment was carried out using a 1 m long LPF and a signal wavelength of 1042 nm with a variable input power. As can be clearly seen in the plot, the TMI threshold sinks very rapidly for signal seed powers below 3 W, which agrees very well with the predictions of the model (blue solid line in Fig. 10b). According to that simulation model this phenomenon is related to an increase of the PD-induced heat-load generated by the absorption of pump photons. The reason is that in short fibers with large cores, such as the LPF used in the experiments, it is not possible to obtain efficient amplification of weak seed signals. Thus, trying to amplify a signal with low average power to a certain output power requires a significant increase of the pump power (compared to the case in which the seed is stronger). Thereby both the inversion and the PD-losses are increased, which leads to the absorption of a higher number of pump photons and, with it, to a higher heat-load that can potentially decrease the TMI threshold [74]. However, in order to contribute to TMI, this PD-induced heat-load created by the absorption of pump photons has to mimic and follow the modal interference pattern that gives rise to the thermally-induced refractive-index grating. Using a beam propagation simulation, very recently it was revealed that this heat-load can indeed strengthen the thermally-induced index grating responsible for TMI [74]. At this point, however, it should be mentioned that, even though these simulations suggests that the reduction of the TMI threshold is mostly due to PD, there is another effect that has been shown to play an important role in determining the TMI threshold for weak seed signals (and which has not been incorporated in the model described in [68]): gain saturation. In fact, it has been widely reported that a lower level of gain saturation in a fiber, as happens when injecting a low power seed, leads to lower TMI thresholds [75,76]. Interestingly, both these effects, the increased PD-induced heat-load and the lower gain saturation, coincide in an active fiber when seeding it with low signal powers and both should lead to a reduction of the TMI threshold. Thus, trying to isolate the individual contribution of each effect, to determine the dominant one, is a difficult task. Quite likely, though, the model presented in [68], overestimates the PD-contribution to the heat-load in situations of low saturation (which compensates for the negative impact of low gain saturation not included in the model).

One important observation that could be extracted from the model presented in [68] is that, for a given few-mode (i.e. multimode) fiber amplifier, usually the TMI threshold is reached for a nearly constant average heat-load [68]. For example, the TMI thresholds presented in Fig. 10a have been reached in good approximation for an average heat-load of 34 W/m at all signal wavelengths. The same is also true for the TMI thresholds plotted in Fig. 10b, which correspond to an average heat-load of ~30 W/m in the fiber for all seed powers. The physical reason that explains this observation is the fact that normally the strength of the thermally-induced index grating (which is intimately related to the TMI threshold) depends almost linearly on the average heat-load in the active fiber [43,68]. This is also a behavior that has been observed when studying TMI in Raman amplifiers [32]. As such, a deviation from this behavior is only expected to be caused by a strong change in the gain saturation level of the fiber [75]. Remarkably, the comparison of several experiments with different Yb-doped fiber laser systems has shown that the TMI threshold is often reached for an approximate average heat-load of 34 W/m [68]. At the light of these observations, the value of an average heat-load of 30-34 W/m has been proposed as a first order estimate to calculate the TMI threshold in various high-power fiber laser systems. It should be said, however, that this rough estimate is controversial due to a lack of generality, since it has only been extensively tested and calibrated with ~1 m long "large pitch" fibers (LPFs) [68]. This means that this estimate works for gain saturation levels typically found in the high power/low gain booster fiber amplifier stages used in high-energy pulsed systems. Since the model does not include the impact of gain saturation on the transverse profile of the thermally-induced index grating, its predictions will lose in accuracy if the saturation level of the simulated amplifier significantly departs from the ones used in the calibration. Additionally, strictly speaking, this estimate is only valid for fibers that are able to guide HOMs even at low power. In fact, it has been observed that for single-mode fibers the TMI threshold is usually reached at the point when the peak heat-load is high enough to allow for the propagation of at least one HOM, i.e. when the fiber becomes multimode due to the thermal profile [17,77,78].

In general, it has been shown that, for small core fibers (i.e. core diameter <40μm), usually the longer the fiber and the lower the V-parameter the higher the TMI threshold [17,77,79,80]. In other words, for this type of fibers the ability to maintain strict single-mode operation at high-average powers leads to higher TMI thresholds. Additionally, it has also been shown that the modal excitation at the seed end of the fiber also has an impact on the TMI threshold [36,77,81]. Moreover, both the pump direction [17,74,80], the seed power [68] and the pump and signal wavelength [65,68,82] seem to influence the TMI threshold. Finally, even though, to the best of our knowledge, no systematic measurements have been published to date, some evidence suggests that the TMI threshold of narrow linewidth amplifiers [83,84] tends to be lower than that of amplifiers emitting broader spectral signals. In this context, some theoretical studies considering sub-MHz linewidths arrive at the conclusion that the bandwidth should not play any significant role in determining the TMI-threshold of narrowband fiber lasers [40]. On the other hand, studies considering broadband fiber lasers (with bandwidths from ~1 nm and upwards) show that the intermodal walk-off can lead to an increase of the TMI threshold [85,86]. Many of these dependencies of TMI have been exploited to develop mitigation strategies for this harming effect and will be described in more detail in section 6.

In any case, what these studies have collectively revealed is that the TMI threshold is not a characteristic value of a fiber design (as sometimes implicitly assumed), but it is rather determined by the whole amplification system. Even though the bulk of the research of TMI has been carried out in fiber amplifiers, this effect has also been explored in other systems. For example, TMI has been investigated in fiber oscillators [58,87,88], and the general conclusion seems to be that the dynamic behavior is somewhat more complex than in amplifiers [58] and the TMI threshold tends to be lower [87]. Likewise, there has been recently some work on double-pass amplifiers, where quasi-static and dynamic mode deformations have been observed [89].

On a different account, another interesting observation related to photodarkening is that the beam of some fiber amplifiers degrades in a time scale of minutes to hours. However, it has to be stressed that this degradation is quasi-static and no beam fluctuations can be observed in this process. This behavior has been explained in [90] as the result of the inscription of a quasi-permanent index-grating in the fiber due to the interplay of the modal interference pattern and photodarkening. It is worth mentioning that a similar quasi-static degradation of the beam profile has been predicted and observed in double-pass amplifiers [89,91,92] and two-core fibers [93,94]. In this case, however, the underlying physics of this effect are slightly different since they do not rely on photodarkening but on the grating created by counter-propagating modes or by the interaction of super-modes.

Even though these effects, strictly speaking, are not TMI, they are so closely related to it, that it can be almost considered as their quasi-static counterpart.

## 5. The importance of the phase shift

As mentioned above, there are two requirements for TMI: the appearance of a refractive-index grating (RIG) along the fiber (in most cases thermally-induced by the modal interference pattern) and the onset of a phase shift between this RIG and the modal interference pattern (MIP) [35]. Actually, as illustrated in Fig. 11, the sign of the phase shift determines the direction of the modal energy transfer: if it is positive, the energy flows from the HOMs to the FM; whereas if it is negative, the energy flows from the FM to the HOMs. In fact, during the beam fluctuations characteristic of TMI, this phase shift does not remain constant but changes in magnitude and sign with time, as can be seen e.g. in the simulations presented in [27]. This means that, during TMI, the energy flow changes direction between the FM and the HOMs, as already shown in Fig. 4.

The schematic diagram shown in Fig. 11 illustrates the three different types of phase shift between the MIP and the RIG: no phase shift (left panel), positive phase shift (central panel), and negative phase shift (right panel). This plot only shows a short 5 cm long section of a high-power fiber amplifier, in which both the MIP and the radially anti-symmetric part of the RIG can be observed. For easy reference, two vertical lines, one black and one white, have been included in the plots. The white line shows the position (along the fiber) of an intensity maximum of the MIP, whereas the black line shows the position of the corresponding maximum of the refractive-index change that forms the RIG. If these two lines fall at the same position along the fiber, i.e. if all the maxima and minima of the MIP and RIG are aligned, there is no phase shift and no energy exchange between the fiber modes can take place (left panel). If the MIP (or a section of it) is shifted towards the fiber input (with respect to the RIG), i.e. if the white line is on the left of the black one, then there is a positive phase shift and the energy can flow from the HOMs into the FM (depending of the strength of the RIG) (central panel).

Alternatively, if the MIP (or a section thereof) is shifted towards the fiber output end (i.e. towards the right) with respect to the RIG, then there is a negative phase shift that allows the energy flowing from the FM towards the HOMs (right panel).

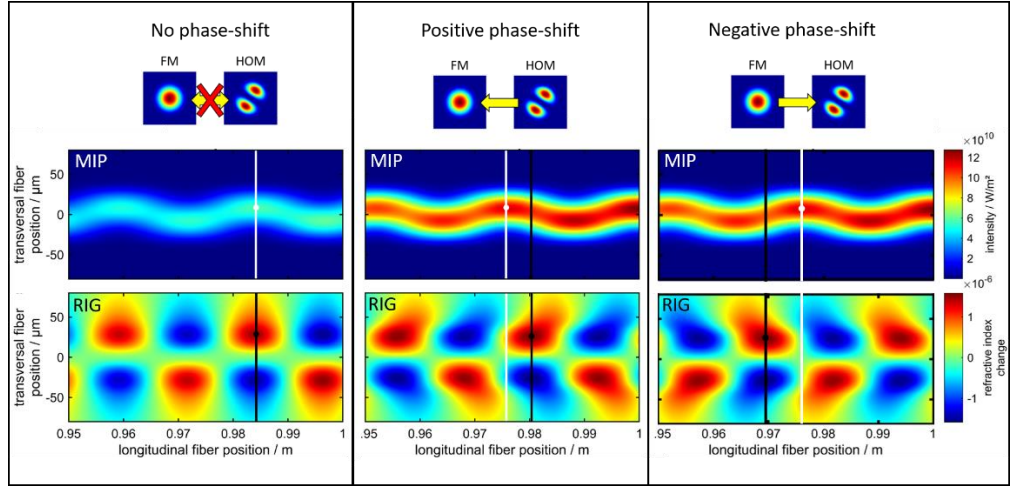

Fig. 11. Schematic representation of the phase shift between the modal interference pattern (MIP) and the thermally-induced refractive-index grating (RIG) in a 5 cm long section of a high-power fiber amplifier (which seed side is on the left-hand side and its output on the right-hand side). The white line indicates the position of a maximum of the MIP, whereas the black line represents the position of the corresponding maximum of the RIG. If these two lines are aligned, there is no phase shift and no energy can be exchanged between the fiber modes (left panel). If the white line (i.e. the MIP) is shifted towards the fiber input side (with respect to the black line/RIG), then there is a positive phase shift and the energy can flow from the HOMs towards the FM (central panel). On the contrary, if the white line (i.e. the MIP) is shifted towards the fiber output (with respect to the black line/RIG), then there is a negative phase shift and the energy can flow from the FM towards the HOMs (right panel). Adapted from Opt. Express 26, 19489 (2018) [95].

A recent published work [95] shows that a change in the pump power can generate a phase shift between the MIP and the RIG. The reason is that, as the pump power changes, so does the output power and, therefore, the heat-load in the fiber. Thus, as the temperature changes, the refractive-index profile in the transversal direction is also modified, as schematically illustrated in Fig. 12. This way, higher temperatures in the fiber (under normal operating conditions) lead to stronger temperature gradients in the transverse direction, which, through the thermo-optic effect, lead to steeper quasi-parabolic changes of the transverse refractive-index profile in the fiber core region [42,96]. This usually results in a separation of the effective refractive indexes ($n_{eff}$) of the FM and the HOM and, thus, it leads to an increase of the numerical aperture of the fiber core, which results in a separation of the effective refractive indexes ($n_{eff}$) of the FM and HOM and might, even, transform a single-mode fiber into a multimode waveguide given enough heat-load [96]. Thus, since the period of the MIP (i.e. the modal beat length) is inversely proportional to the $n_{eff}$ difference between the FM and the HOM [95], a change in the pump power will lead to a stretching or compression of the MIP. Whereas the MIP, which is nothing but the result of an interference, can react instantaneously to a change in the temperature gradient in the fiber, the RIG cannot since it has to evolve from a previous state. This means that, whenever there is a change in the temperature gradient of the fiber, the MIP will react instantaneously and the RIG will lag behind, thus giving rise to a phase shift similar to those shown in Fig. 11.

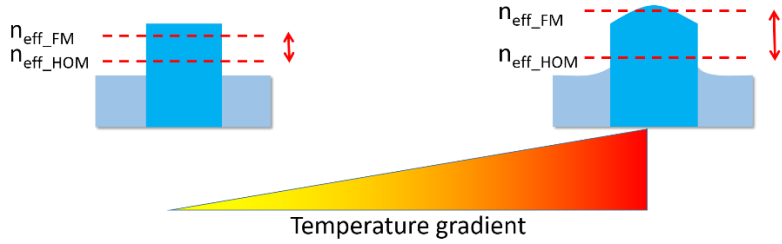

Fig. 12. Schematic representation of the thermally-induced changes in the refractive-index profile of an optical fiber. As the temperature increases, so does the temperature gradient, which leads to the onset of a steeper and steeper nearly parabolic refractive-index change in the fiber core. This, in turn, results in a stronger separation of the effective refractive indexes ($n_{eff}$) of the FM and HOM.

The knowledge gained from [95] opens the door to probing the two long-standing theories for the trigger of TMI described in section 3. In summary, the first of these theories says that the transverse modes in the fiber have slightly different central frequencies [35], which leads to a travelling MIP. This results in the presence of a strong phase shift between the MIP and the RIG even at low average powers. Therefore, in the frame of this theory, TMI is triggered when the RIG reaches a certain strength. The other theory says that the transverse modes are all excited with the same frequency content, which leads to a quasi-static RIG [34,38]. This RIG becomes stronger (i.e. the thermally-induced refractive-index changes become deeper) as the average power extracted from the fiber amplifier increases. The stronger the RIG, the more sensitive it becomes to external perturbations that lead to small phase shifts (e.g. pump-power noise). In other words, with very strong RIGs even very small phase shifts suffice to cause a significant energy exchange between transverse modes of the fiber. Thus, in the frame of this theory, TMI are triggered when the sensitivity of the system to noise-induced phase shifts becomes high enough.

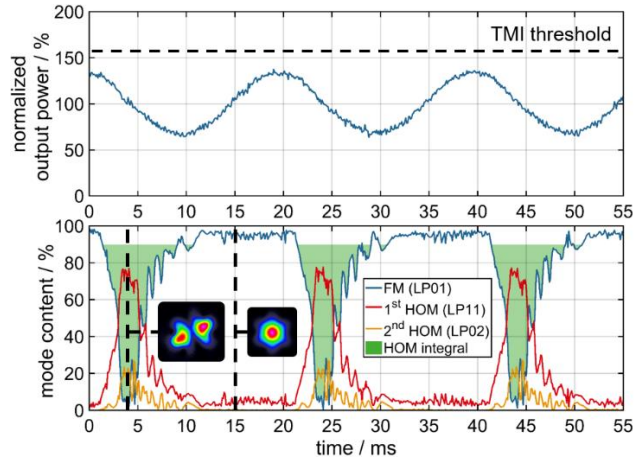

Fig. 13. Evolution of the normalized output power (average power 150 W) with respect to the TMI threshold of the system (upper graph). The pump power was sinusoidally modulated with 50 Hz and a modulation depth of ±50 W. The lower plot shows the evolution of the modal content for three modes: the FM, the first HOM (i.e. the $LP_{11}$) and the second HOM (i.e. the $LP_{02}$). The region of strong energy transfer into the HOMs is shaded green. Adapted from Light Sci. Appl. 7, 59 (2018) [43].

The way that these two theories can be probed is as follows: by modulating the pump power of the amplifier it will be possible to induce periodic changes in the phase shift between the

MIP and the RIG (according to [95]). The amplitude of these phase shifts can be controlled by the frequency and depth of the pump modulation as shown in [43]. Thus, if the first theory is correct, the induced phase shifts will only lead to a significant modal energy transfer at average powers close to the TMI threshold, whereas if the second theory is correct, a significant modal energy transfer should be observed at powers well below the TMI threshold.

The experiments described in [43] were carried out in a Yb-doped LPF amplifier of ~1.1 m length with an active core of ~65 µm and a TMI threshold of 233 W (measured according to [22]) that was seeded with 5 W at 1030 nm and counter-pumped at 976 nm. Hereby the pump power was modulated with frequencies ranging from several Hz up to 2.5 kHz and the output beam was recorded with a high-speed camera able to capture up to 67000 frames per second. The frames were later (i.e. offline) analyzed by an algorithm that is able to reconstruct the amplitude and phase of the different transverse modes of the fiber from the intensity images captured in each frame [23]. One of the results from these experiments can be seen in Fig. 13. Here the system was operated at an average power of 150 W (i.e. significantly below the TMI threshold) and the pump was modulated with a frequency of 50 Hz and an amplitude of ±50 W. This was enough to keep the output average power below the TMI threshold at all times, as can be seen in the upper plot of Fig. 13. In spite of this, as revealed in the lower plot of Fig. 13, the modal content analysis shows that there is a very strong energy transfer between the FM and the HOMs of the fiber. In fact, as can be seen, the HOM content grows in the falling edge of the output signal power, i.e. when the temperature of the fiber sinks and the MIP is stretched or, in other words, when a negative phase shift is induced. In this region, the beam profile acquires the shape of a $LP_{11}$ mode. On the contrary, the HOM content drops a soon as the inflection point of the sinusoidal modulation is reached, because the negative phase shift decreases in magnitude. The HOM content reaches its minimum in the rising edge of the sinusoidal modulation, where the temperature increases and the MIP gets compressed or, in other words, when a positive phase shift is induced. In this region, the beam has the intensity profile of the FM. This behavior could be well reproduced with the model described in [95].

The fact that, when inducing a phase shift between the MIP and the RIG, there is a strong energy transfer at average powers significantly below the TMI threshold (in [43] it is reported that such energy transfer has been observed even at powers < 100 W), implies that the RIG is strong enough to allow for modal energy transfer already at relatively low average powers. This experimental evidence seems to favor the second of the theories for the trigger of TMI.

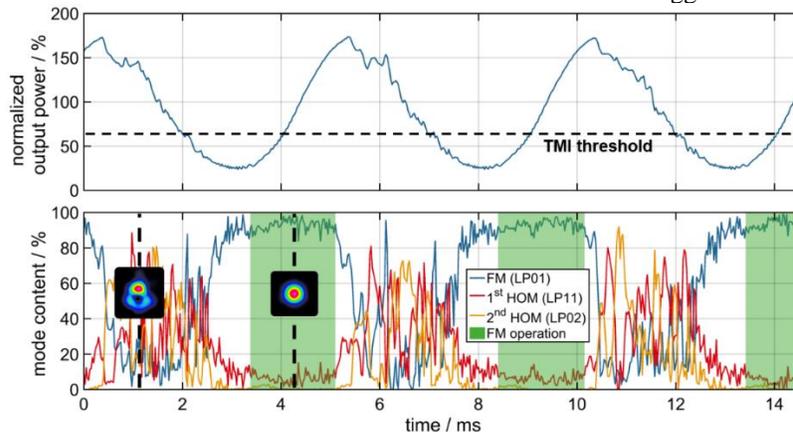

Fig. 14. Evolution of the normalized output power (average power 350 W) with respect to the TMI threshold of the system (upper graph). The pump power was sinusoidally modulated with 200 Hz and a modulation depth of ±262.5 W. The lower plot shows the evolution of the modal content for three modes: the FM, the first HOM (i.e. the $LP_{11}$) and the second HOM (i.e. the $LP_{02}$). The region of strong energy transfer into the FM is shaded green. Adapted from Light Sci. Appl. 7, 59 (2018) [43].

Another crucial observation presented in [43] is shown in Fig. 14. In this case the system was operated beyond the TMI threshold for most of the time. In fact, the output average power was 350 W, i.e. around 1.5 times higher than the TMI threshold. As can be seen, while in the falling edges of the power modulation the modal content fluctuates very strongly between the different transverse modes of the fiber, the percentage of energy contained in the FM increases towards 100 % in each rising edge of the power modulation. Actually, in this period of time the beam profile resembles that of the FM. This happens in spite of the system being operated at peak output powers that are well above twice the TMI threshold. Such a behavior points towards a control of the modal energy transfer which, in the rising edge of the modulation, lets the energy flow into the FM, thus resulting in a beam cleaning.

This is a very important observation since it demonstrates the feasibility of developing a completely new family of TMI mitigation strategies [97] based on the control of the phase shift between the MIP and the RIG. This will be discussed in section 6.

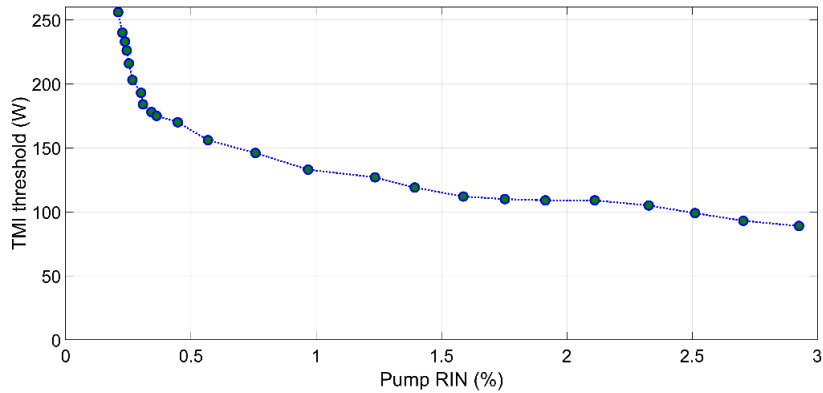

Fig. 15. Evolution of the TMI threshold as a function of the relative intensity noise (RIN) of the pump. Adapted from PhotoniX, 1 (2020) [44].

There have been further theoretical [53] and experimental [44] studies related to the phase shift. Thus, for example, taking into account that the pump modulation can lead to a phase shift and, therefore, to modal energy transfer even below the threshold, it has been speculated that pump intensity-noise could be the ultimate driver for TMI. Even though some theoretical work on the impact of pump/seed modulation has been done some time ago [40,45,46,67,98], the first systematic measurements of the impact of the relative intensity noise (RIN) of the pump have been presented very recently [99]. Herein white RIN (between 0 Hz and 2 kHz) has been applied to the laser diode that pumps a 1.1 m long LPF with ~65 µm core diameter from the counter-propagating direction. The system was seeded with 5 W at 1030 nm and the TMI threshold of the free-running system was 256 W. These experiments have demonstrated that an increase of the pump RIN leads to a significant reduction of the TMI threshold. In particular, as shown in Fig. 15, for this system the TMI threshold dropped by ~63 % when the pump RIN reached a value of ~3 % [44,99]. Additionally, contrary to some of the theoretical predictions, it has also been shown that the impact of the seed intensity-noise in saturated amplifiers is significantly weaker than that of the pump intensity-noise [44]. In fact, in these experiments, the TMI threshold dropped by just ~13 % for a seed RIN of ~3 %, whereas the maximum observed drop of the TMI threshold was ~40% for ~22% seed RIN. These measurements strengthen the view that pump intensity-noise could play a big role in driving TMI in high-power fiber laser systems.

## 6. Mitigation of transverse mode instability

Over the years, researcher all around the world have been proposing ways to increase the TMI threshold. Several of these methods and strategies will be presented and discussed in this section.

As it has been mentioned in section 3, there are two conditions necessary to obtain the beam fluctuations characteristic of TMI: the appearance of a (thermally-induced) transient index grating written by the modal interference pattern and a phase shift between them. Should any of these two conditions not be fulfilled, then the beam would remain stable. Thus, it is no surprise to find out that the mitigation strategies can be divided in two big families: the ones acting on the RIG and the ones acting on the phase shift. To date, most of the TMI mitigation strategies try to reduce the strength of the RIG but, recently, some methods have been proposed which are based on manipulating the phase shift. In this section representatives of both types of strategies will be presented and discussed.

Independently of the family they belong to, TMI mitigation strategies can be classified in two general categories: passive and active [100]. The passive methods are those that are not controlled by any external element, whereas the active methods involve an external control mechanism.

### *6.1 Passive mitigation strategies*

#### 6.1.1 Fiber design

As already explained in previous sections, the ultimate cause of TMI is the fact that active fibers in high-power operation are able to guide several transverse modes. Such a simple statement highlights the fact that the fiber design can play a decisive role in increasing the TMI threshold.

A prominent parameter of any fiber design is the core size. Interestingly, there is currently a heated debate about the impact of such a parameter on the TMI threshold. Whereas some authors defend that the TMI threshold should decrease with the square of the core radius [5,46,54], others maintain that the dependence of the TMI threshold on the core size should be weak if parameters such as the numerical aperture or the V-parameter are left constant [73,75,85,101,102]. The underlying problem is that it is extremely difficult to carry out a systematic study since, in practice, changing the core dimensions tends to change the guiding properties of the fiber, which reduces the comparability of the results. Moreover, since the TMI threshold is dependent on many different parameters of the fiber (e.g. NA, fiber length, clad-to-core ratio) even doing a literature research and plotting the experimentally measured TMI thresholds against the core size, as done in [5], is not necessarily trustworthy since widely different fibers are compared with each other (different lengths, different NA, different signal and pump wavelengths, different fiber designs, etc.). Thus, this remains a point that needs further research to be fully clarified.

On the other hand, everybody agrees that it is particularly desirable to use a fiber design that is able to keep single transverse-mode operation even at high average-powers to increase the TMI threshold. This is the successful strategy followed in many recent CW high-power fiber laser systems [17,80]. The challenge hereby is that fibers for high-power operation, as already mentioned in section 1, usually have large cores to mitigate non-linear effects, which hinders single transverse-mode operation. There are several reasons for this: on the one hand, the numerical aperture (NA) of the fiber core cannot be arbitrarily reduced due to technical and physical limitations. In spite of this, few mode fibers with low V-parameters can still be forced into single-mode operation by bending them with small radii, which results in high losses for the HOMs. This approach has been shown to increase the TMI threshold by >50 % [24,67,79,103]. However, even fibers that were originally single-mode start supporting the propagation of several transverse modes at high powers due to the development of a thermally-induced parabolic index change in the fiber core [96,104]. Therefore, it can be concluded that active fibers will usually guide several modes in high-power operation (this is particularly true for the fibers used in ultrafast systems). That is why large-core fibers are specially designed to favor the propagation and amplification of the desired FM. With this strategy it can be achieved that the FM becomes the dominant mode at the end of the active fiber, in which case it can be spoken of "effective single-mode" operation.

Several strategies to achieve effective single-mode operation have been developed in the last decade. These fiber designs exploit different principles to discriminate between the FM and the HOMs such as, for example, the introduction of high propagation losses for the HOMs [105], the creation of a photonic bandgap [106], selective coupling of the HOMs to neighboring cores [13,14], or the delocalization of HOMs [12,107,108]. Whereas fibers based on the first two concepts try to extinguish the HOM by generating very high propagation losses for them, fibers based on the last two approaches try to selectively remove the HOM from the active region. This last strategy (i.e. the selective removal of the HOMs) is very promising for mitigating TMI and deserves a closer inspection. In this context, in the following we will analyze the principle of HOM delocalization since this approach has established itself as one of the most successful ones for fibers with MFD > 50 µm. The first very large mode area fiber design that incorporated HOM delocalization was the "large pitch" fiber (LPF) [107], but several new designs exploiting this principle have been proposed afterwards [13,15,109]. In any case, regardless of the particular fiber design exploiting it, the concept of delocalization refers to the fact that the HOMs are "pushed away" from the core, as shown in Fig. 16, which can be achieved with a careful design of the fiber structure. Hereby the effect of "avoided crossings" for the HOMs is exploited [110] in order to achieve that the HOMs have a significantly lower overlap with the doped region than the FM. This results in a twofold advantage: on the one hand the HOMs are not as efficiently amplified as the FM and, on the other hand, they are not easily excited by a Gaussian seed beam launched into the fiber. In fact, the combination of these two effects results in a significantly lower portion of the signal power being guided in the HOMs than in the FM at the end of the fiber [12]. Furthermore, it has been recently demonstrated that asymmetric/aperiodic structures [12,109,111,112] can enhance the delocalization of HOMs beyond what is possible with the basic LPF design, as shown in Fig. 16.

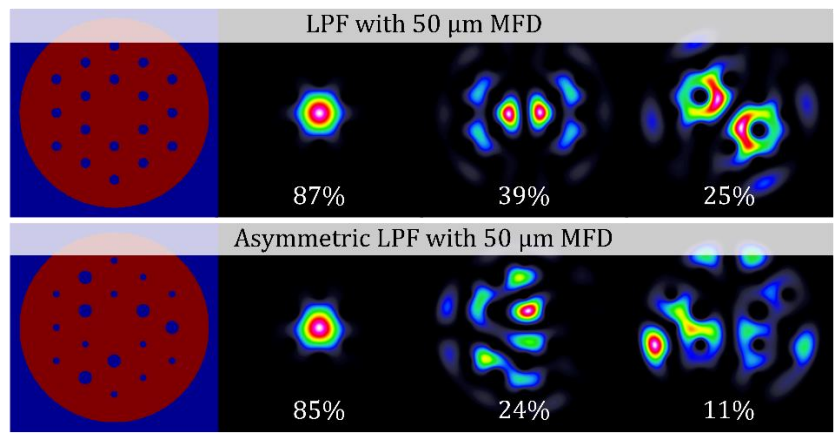

Fig. 16. Example of two large-mode-area fibers that exploit the delocalization of HOMs. Above: large pitch fiber (LPF). Below: asymmetric LPF with increased delocalization. Under each mode there is its overlap with the doped region expressed as a percentage of the energy. Adapted from Optica 1, 233 (2014) [12].

At this point it is important to realize that modal delocalization has a positive effect on the TMI threshold. In fact, fibers exploiting mode delocalization show a TMI threshold that is 2-3 times higher than that seen in conventional fiber designs of similar dimensions [113]. The reason is that this approach is able to weaken the RIG since the HOM content at the beginning of the fiber is lower and the HOMs are not so efficiently amplified along the fiber as the FM (thus, the relative power content of the HOMs is reduced upon amplification).

As already mentioned, the delocalization results in an overlap of the HOMs with the doped region that is lower than that of the FM. A similar effect can also be achieved if the doped region is made smaller than the actual core size [114,115]. However, in contrast to fibers that exploit delocalization, the excitability of the HOMs is not reduced in the designs with reduced doped region. This strategy is called "preferential gain" or "confined doping" and it has been tested in different fiber designs [45,116–118]. Thus, the positive impact of this approach has been demonstrated both theoretically [45,46,75,118] and experimentally [116,117]. For example, using a rod-type fiber it was shown that the TMI threshold can be increased by a factor of 2-3 with respect to that obtained with a fiber without preferential gain, albeit with a 50 % longer fiber in the preferential gain case [116].

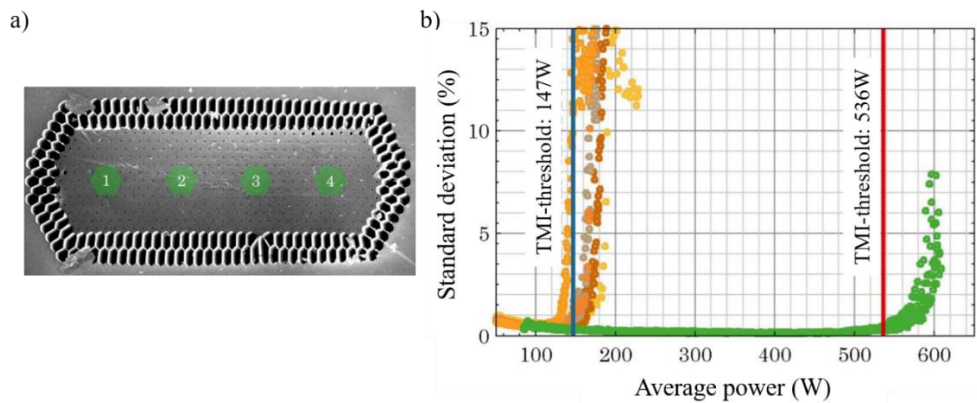

Fig. 17. a) Microscope image of a multicore fiber with 4 Yb-doped cores and b) the measurement of the TMI threshold for each individual core (yellow-brown points) results in an average value of 147 W. After the combination of the power of all the cores into a single beam the TMI threshold is reached at a total average power of 536 W. Adapted from Opt. Lett. 39, 2680 (2014) [119].

An interesting possibility not to solve, but to get around the problem of TMI is to use multicore fibers [119,120], which, as their name indicate and as shown in Fig. 17a, have several signal cores within a single pump cladding. However, it is important to pay attention to the design of the multicore fiber. Hereby, if the cores are very close together, as it used to be the case in the early designs for high-power operation, they will be optically (and thermally) coupled with each other. This can result in lower TMI-thresholds (or, at least, in no improvement) than in single-core fibers (~100 W in the experiment presented in [120] which used a 1.5 m long, 18-core photonic crystal fiber with coupled cores of ~12 µm diameter). On the other hand, with a careful design of the fiber it is possible to have the cores both optically and thermally isolated from each other, so that they operate as virtually independent waveguides [119]. If this condition is fulfilled, then the TMI threshold of each individual core will be similar to that of the single core fiber (around 147 W in the example shown in Fig. 17b) but the combination of the beams emitted by them into a single output beam will result in an increase of the overall TMI threshold by a factor roughly equal to the number of cores (536 W in the example shown in Fig. 17b, which employed a 1.15 m long, 4-core photonic crystal fiber with independent cores of 50 µm in diameter). Recently, a high-power 16-core fiber has been demonstrated [121] and it is expected that such fibers will enable multi-kW ultrafast fiber laser systems in the near future. In any case, the phenomenon of TMI in multicore fibers remains fairly unexplored up to now with only a few experimental observations [119,120] and theoretical studies [93,94] having been reported so far.

6.1.2 Fiber core composition

This section describes the optimization of the fiber core material from the point of view of TMI. It should be mentioned, however, that some of the optimization studies that will be presented in this and the next subsection are based on the experimental observation that the TMI threshold is usually reached by an average heat-load of approximately 34 W/m in Yb-doped (multimode) fibers, as already discussed in section 4. Thus, the simplified model presented in [68] has been used, which is not widely accepted in the community and, therefore, it is important to briefly discuss the validity of this model. The model has been calibrated for gain saturation levels typically found in the high power/low gain booster fiber amplifier stages found in high energy pulsed systems. In fact, since the model does not include the impact of gain saturation on the transverse profile of the thermally-induced index grating, its predictions will lose in accuracy if the saturation level of the simulated amplifier significantly departs from the ones used in the calibration. Thus, the model has been extensively tested and calibrated with ~1 m long "large pitch" fibers (LPFs) [68], and even though it also seems to cast good results with other fiber designs and lengths [74], most of the optimization studies have been done simulating fiber amplifying systems based on LPFs to minimize the impact of gain saturation. In spite of this, the general guidelines and trends shown in the following for LPFs should be equally applicable to other fiber designs (although the resulting improvement in terms of TMI threshold might be different to the one presented here).

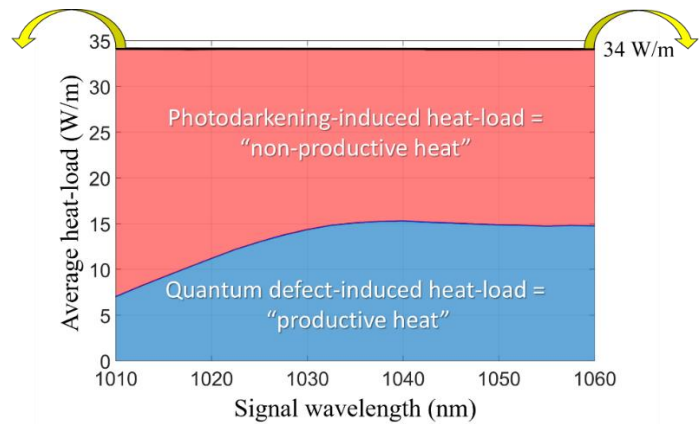

Fig. 18. Interpretation of a fiber laser systems as a heat-load bucket with a limited capacity (34 W/m). When the bucket overflows (yellow arrows), then TMI appears. Reprinted from Opt. Express 23, 20203 (2015) [68].

The (controversial) finding that the TMI threshold is reached in good approximation for a constant average heat-load (with the constraints mentioned in section 4) allows for a useful interpretation of a fiber laser system from a thermal point of view. Thus, in this interpretation a fiber system can be understood as a heat-load bucket with a limited capacity, as shown in Fig. 18 (which values correspond to the simulations of the experiment presented in Fig. 10a). This heat-load bucket (that has a capacity of 34 W/m in the example for the LPF) can be filled with any kind of heat, i.e. no matter whether it is "productive" (i.e. QD-induced heat) or "un-productive" (e.g. PD-induced heat). Independently on the way in which the heat-bucket is filled, once that it is full and overflows, the TMI threshold is reached. From this interpretation it becomes clear that, in order to increase the TMI threshold of a fiber laser system, it is necessary to fill the heat-bucket with as much productive heat as possible. Unfortunately, as can be seen in Fig. 18, the productive heat in the example (of an LPF) not even fills the heat-bucket to half of its capacity, which leaves a lot of room for improvement.

In section 4 it was already discussed in detail that the PD-losses have a strong impact on the TMI threshold. Crucially, since these losses depend on the $Yb^{3+}$-ion concentration and on the overlap of the modes with the doped region, they can be reduced by a careful optimization of the fiber core composition.

There are several degrees of freedom when designing a fiber core doped with laser-active ions and these are: the doping concentration, the doped area and the distribution of the active ions. The first of these parameters, the doping concentration, is also the most important one since it determines the maximum amplification and PD-losses. Normally a high doping concentration of $Yb^{3+}$-ions is desirable because it results in a higher amplification/stored energy in the fiber. However, the quadratic dependence of the PD-losses on this parameter [62] implies that higher $Yb^{3+}$-ion concentrations will result in lower TMI thresholds. Note that without PD this parameter alone should have no big influence of the TMI threshold in Yb-doped fibers [122], but it could still be important in Tm-doped fibers [31]. Therefore, it is mandatory to find a compromise between a high amplification and low PD-losses, which is to use the lowest possible concentration that still leads to the desired laser performance. The next important parameter is the doped area which ultimately influences the TMI threshold through the overlap of the modes with the doped region. This has already been discussed in section 6.1.1 and it is, therefore, only mentioned at this point.

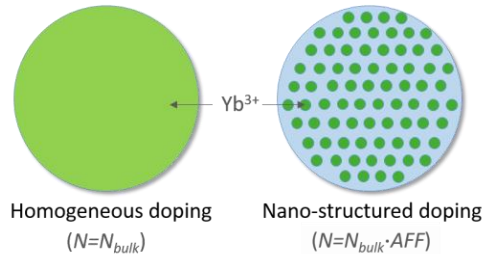

Fig. 19. Two different possibilities in which the $Yb^{3+}$-ions can be distributed in the fiber core. (Left) The homogeneous distribution is typical in step-index fibers that have been fabricated with the MCVD [123] or the REPUSIL [124] preform manufacturing processes. (Right) In photonic crystal fibers the $Yb^{3+}$-ions are usually concentrated in isolated doped islands that are smaller than the wavelength of the light. This distribution is called nano-structuring. Reprinted from Opt. Express 23, 20203 (2015) [68].

The last degree of freedom is, as illustrated in Fig. 19, the distribution of the $Yb^{3+}$-ions across the doped area. This distribution depends on the manufacturing process of the fiber. For example, in step-index fibers, which preform has been fabricated using either the MCVD [123] or the REPUSIL [124] process, the $Yb^{3+}$-ions are usually homogeneously distributed in the doped region (see left-hand side of Fig. 19). In contrast, photonic-crystal fibers are usually fabricated using the "stack and draw" process, which results in the $Yb^{3+}$-ions being concentrated in isolated doped islands that are smaller than the wavelength of the light (see right-hand side of Fig. 19). Since between these isolated doped islands there is undoped glass, the light sees an effective $Yb^{3+}$-ion concentration ($N$), which is the average across the complete doped region. The effective $Yb^{3+}$-ion concentration ($N$) is smaller than the local ion concentration in each of the doped islands ($N_{bulk}$) by a factor called "area-filling fraction" (AFF) [68]. The AFF is the ratio of the sum of the areas of all the doped islands to the total area of the doped region. Thus, the AFF in a passive fiber is 0 and 1 in a step-index fiber fabricated by the MCVD process (left-hand side of Fig. 19). Due to the non-linear dependence of PD on $N_{bulk}$ already discussed in section 4, the lower the AFF the higher the PD losses for any given effective $Yb^{3+}$-ion concentration ($N$). Therefore, since the AFF determines the overlap of the modes with the doped region, it can be expected that this parameter will also influence the TMI threshold.

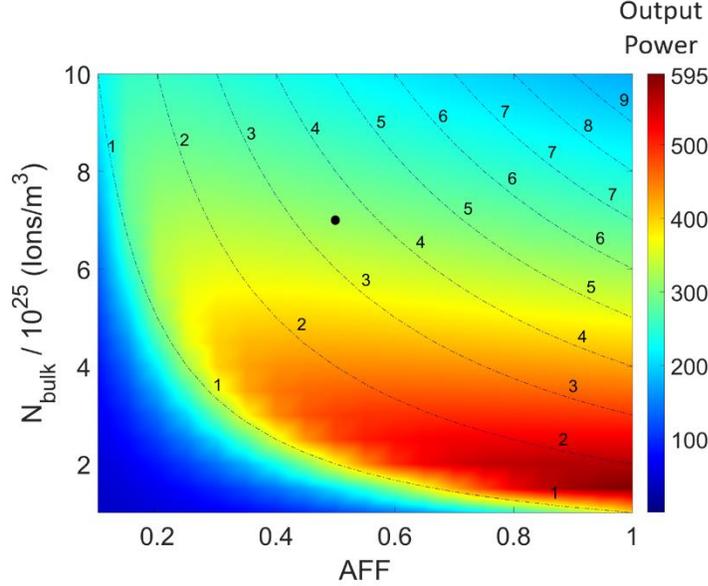

Fig. 20. Dependence of the TMI threshold of a 1.2 m long LPF on the $Yb^{3+}$-ion concentration ($N_{bulk}$) and the area filling fraction of the core (AFF). The isolines represent combinations of $N_{bulk}$ and AFF that result in a constant effective $Yb^{3+}$-ion concentration ($N$) and, therefore, in constant fiber dimensions since the pump absorption has been set constant at 24 dB/m for all points. The black dot shows the typical parameters used in a conventional LPF. Adapted from Opt. Express 24, 7879 (2016) [74].

The dependence of the TMI threshold on the $Yb^{3+}$-ion concentration ($N_{bulk}$) and the AFF factor has been studied in [74] using the model described in [68]. Independently and almost simultaneously a similar study, using a different model, has been published in [125]. Even though the fiber geometry simulated in these two papers is different and, therefore, the direct comparability of the results is not straightforward, both works seem to have a good qualitative agreement.

Fig. 20 exemplarily shows the simulations presented in [74], which were carried out for an LPF with 80 µm core diameter (V-parameter of 7) and a length of 1.2 m. In spite of the different combinations of AFF and $N_{bulk}$ parameters used in the simulation, the pump absorption of the fiber was always left constant (~24 dB/m for a pump at 976 nm), which implies that the pump cladding diameter had to be correspondingly adjusted in every case. The result from this investigation can be seen in Fig. 20, where it can be observed that, even though the TMI threshold shows the strongest dependence on the value of $N_{bulk}$, its change with AFF cannot be neglected either. In general, it seems that for any given effective ion concentration $N$ (represented by the isolines in Fig. 20), the best strategy to obtain the highest possible TMI threshold is to choose a low $N_{bulk}$ (but not lower than ~$2 \cdot 10^{25}$ ions/m$^3$) and the highest possible AFF. In this regard, it can be concluded that the preform manufacturing processes based on MCVD or REPUSIL are advantageous since AFF can be very high. This is the strategy followed by some groups and which has led of the recently published performance of 4.3 kW average-power from a Yb-doped MOPA systems with diffraction-limited beam quality [17].

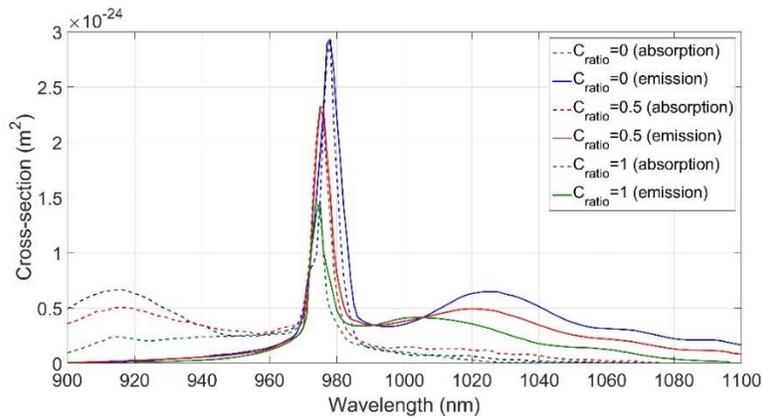

Fig. 21. Absorption (dashed lines) and emission (solid lines) cross-sections of Yb-doped silica for different amounts of Phosphorous added to the glass composition. Reprinted from Opt. Express 26, 7614 (2018) [126].

Another popular way of reducing photodarkening in Yb-doped fibers is to change the composition of the host glass of fibers, as illustrated in Fig. 21. In fact, it has been shown that by adding phosphorous to silica it is possible to significantly reduce the PD losses [127,128]. This is a very attractive approach since it does not compromise the gain and/or energy stored in the fiber (as reducing the Yb-ion concentration does). The main problem with this approach is that, when adding phosphorous to the glass composition the cross-sections of the Yb-ions are reduced, as can be seen in Fig. 21. This implies that, in order to keep the amplification performance constant, it is necessary to increase the Yb-ion concentration in fibers co-doped with phosphorous. However, crucially, as already mentioned before, the PD losses depend quadratically on the ion concentration. This fact can neutralize to a large extent the potential benefits of using a phosphorous co-doped core composition in terms of PD-loss and TMI threshold. This has been addressed in [126], where a thermal analysis of Yb-doped fibers with core compositions based on different ratios of $P_2O_5$ and $Al_2O_3$ has been presented. Hereby, it has been studied at which output average power the different fiber core glass compositions reach a certain average heat-load (in this case, as discussed before, 34 W/m). The idea behind this is that fibers that develop a lower heat-load will have a higher TMI threshold. Thus, even though the exact value of the average heat-load at which the TMI threshold might be reached can differ from the value used in [126], the general dependence of this parameter on the fiber glass composition is still expected to follow the results of the simulation.

One of the fibers simulated in this study is a 1.2 m long LPF with ~80 µm core diameter (which results in ~70 µm mode field diameter), AFF = 0.5, 255 µm pump–cladding diameter and 18 dB/m pump absorption at 976 nm. The fiber is seeded with 30 W at 1030 nm and pumped at 976 nm. Such a fiber has similar parameters to the LPFs used in state-of-the-art ultrafast fiber laser systems. The ratio between the concentration of $P_2O_5$ ($C_{P2O5}$) and $Al_2O_3$ ($C_{Al2O3}$) was characterized using the parameter $C_{ratio} = \frac{C_{P2O5}}{C_{P2O5}+C_{Al2O3}}$ . Hereby, the concentration of phosphorous in the core is increased, which leads to a change of the Yb-ion concentration to keep the amplification efficiency (defined as the ratio between the extracted power and the pump power) at a constant value of 85 %. This leads to a variation of the PD losses and, in turn, of the heat-load in the fiber for different glass compositions. It should be mentioned that this study uses the lowest measured PD loss values for each core composition [127] as a reference. Thus, the evolution of the TMI threshold (here defined simply as the average power at which an average heat-load of 34 W/m is reached) with the core glass composition is shown in Fig. 22. As can be observed, introducing phosphorus in the glass composition is expected to have a positive repercussion in the TMI threshold, which reached an optimum when the concentration of $P_2O_5$ and $Al_2O_3$ is equal ($C_{ratio}$ = 0.5). The positive impact of the inclusion of of $P_2O_5$ in the core composition is expected to be even more pronounced in long fibers with smaller cores, as shown in [126] where the TMI-threshold of a 15m long 20/400 SIF increases from ~3.5 kW for $C_{ratio}$ = 0 to >5 kW for $C_{ratio}$ = 0.5. In fact, a $C_{ratio}$ = 0.5 was the core glass composition of choice in some of the latest demonstrations of high-power CW fiber laser systems with TMI-free operation at ~4.3 kW [17,79].

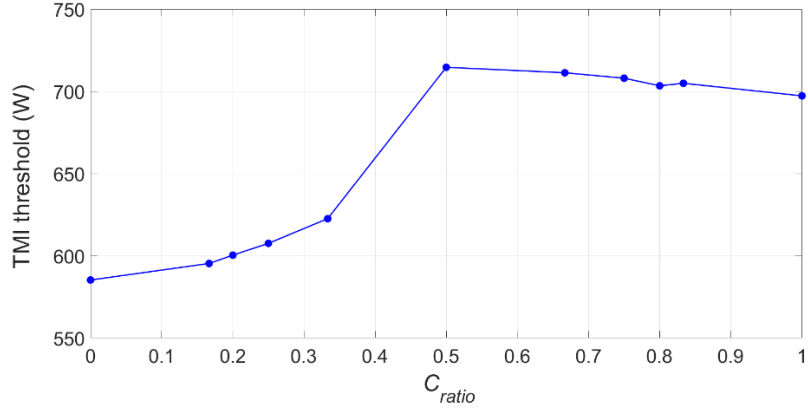

Fig. 22. Expected evolution of the TMI threshold (here defined as the average power at which an average heat-load of 34 W/m is reached) as a function of the core glass composition. Reprinted from Opt. Express 26, 7614 (2018) [126].

Another recent approach that has shown promise to significantly reduce photodarkening in Yb-doped fibers is to use cerium (Ce) as a co-dopant [129]. Unfortunately, no systematic measurements of the impact of Ce on the TMI threshold have been carried out to date, but an increase of the average power of fiber laser systems is expected from this core glass composition.

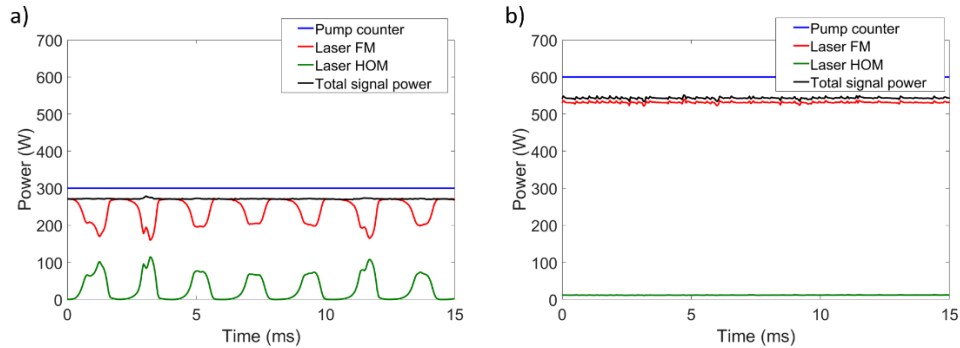

Fig. 23. Comparison of the temporal evolution of the modal content in an 80 µm core, 1 m long fiber: a) made of silica and pumped with 300 W at 976 nm and b) made of YAG and pumped at 600 W at 976nm.

A more radical approach to mitigating TMI is to completely change the host material of the fiber as suggested in [7]. This strategy would mean abandoning silica for materials with higher thermal conductivity (but comparable thermo-optic coefficients) such as, for example, crystalline hosts. One good candidate for this would be YAG, which thermal conductivity is roughly 8 times higher than that of silica and its thermo-optic coefficient is about 20 % lower. Even though from the thermal point of view YAG seems to be an excellent choice, one of its main drawbacks as a fiber material for high-power operation is that it has a non-linear coefficient which is roughly two-times higher than that of silica. Anyway, if this can be tolerated in the system, then YAG should allow for a higher TMI threshold, as shown in Fig. 23.

The simulation results show a comparison between two 1 m long fibers with 80 µm core diameter (V-parameter 7), 228 µm pump core diameter, doped with $3.5 \cdot 10^{25}$ $Yb^{3+}$-ions/$m^3$ but made of two different host materials: silica (Fig. 23a) and YAG (Fig. 23b). The plots in Fig. 23 show the temporal evolution of the modal content for the FM (red line) and for the HOM (green line) together with the evolution of the pump power (blue line) and total signal power (black line). Note that the silica fiber shows mode instability already at a power of ~270 W (Fig. 23a), whereas the YAG fiber shows no modal content fluctuations even at twice the power (Fig. 23b). A recent study predicts that the use of YAG as a host material for fibers could lead to an increase of the TMI threshold by a factor up to 28 [130].This underlines the virtues of changing the host material of the fiber in order to mitigate TMI.

Unfortunately the technology of crystalline fibers is not mature enough to allow for a replacement of silica in high-power fiber laser systems yet. However, encouraging progress is steadily being made in this direction [131–133], which might provide a very promising route to mitigate TMI in the future.

### 6.1.3 Quantum defect and gain saturation

The optimization strategies presented in the previous subsection are aimed at reducing the PD losses as a means to increase the TMI threshold. However, there are other mitigation strategies that would even work without PD in the fiber. Two of these methods to increase the TMI threshold will be discussed in this subsection.

The first of these methods is particularly promising for the case when the QD is the dominant heat source in the fiber. This method pursues a reduction of the QD by decreasing the spectral separation between the pump and the signal wavelength [100,134]. There are at least two possibilities to do this: by shifting the pump towards longer wavelengths while using a constant signal wavelength or by shifting the signal towards shorter wavelengths while using a constant pump wavelength. Any arbitrary combination of these two cases is, obviously, also possible. The main drawback of this strategy is that this shift of the wavelengths can negatively impact the performance of the fiber laser system, in particular regarding the amplification and the extractable energy. Therefore, when choosing this mitigation strategy, some compromises have to be met that might make it more or less attractive depending on the application. In spite of this, this method offers a way to significantly increase the TMI threshold, and it can be exploited when using tandem pumping as it will be discussed in subsection 6.1.4.

Another passive method that can lead to an increase of the TMI threshold is the reduction of the pump absorption or, equivalently, an increase of the gain saturation in the fiber amplifier [75,76,100]. The physical reason for this dependence is that a higher level of gain saturation usually results in stronger transversal hole-burning and, thus, in a transversal inversion profile approaching a flat-top shape, which eventually leads to a more homogeneous transverse thermally-induced index grating and contributes to increase the TMI threshold [75]. This effect also explains the observed dependence of the TMI threshold on the fiber length, since in long fibers the inversion levels are generally lower and more prone to transversal hole burning. In fact, many experimental results published worldwide seem to indicate that the fiber length is one of the parameters with the largest impact on the TMI threshold [68]. In particular, this experimental evidence suggests that longer fibers show higher TMI thresholds. Besides, it has been reported that in long fibers the linewidth of the seed laser can play a role in determining the TMI threshold [85,86]. In fact, the prediction is that in long fibers (i.e. several meters long) the TMI threshold will be higher when using broadband seeds, because after a certain propagation distance the HOM and the FM will lose coherence (or will exhibit a large walk-off) and not interfere with each other anymore. This sets a natural limit to the maximum length of the MIP and the RIG. Thus, the broader the spectral bandwidth of the seed, the shorter the length of the interference pattern and, therefore, the shorter the RIG, which leads to higher TMI thresholds (if the fiber is longer than this maximum RIG length). Even though some experiments have shown a good correlation between an increase of the TMI threshold (by as much as ~50 %) and an increase of the seed linewidth [84], this point, has yet to be experimentally studied in a systematic way.

### 6.1.4 Pump radiation: direction and wavelength

There are two important parameters of the pump that can potentially impact the TMI threshold of a fiber laser system: the pump direction and its wavelength.

The pump direction is also an important design parameter that influences the TMI threshold in a fiber laser system. Most of the theoretical studies done up to now coincide in pointing out that the lowest TMI threshold is usually obtained for the co-propagating pump configuration [46,76,81,135–137]. The reason for this is related to the lower level of gain saturation obtained in the co-pumping configuration. Interestingly, the threshold change with the pump direction seems to become negligible for fibers with a low cladding to core ratio [135], such as those used in high-power ultrafast fiber laser systems.

It has been pointed out that a bi-directional pump configuration might offer some advantages over the counter-propagating one [84,137–139], even though there are contradicting studies on this subject [81]. This most likely means that the most beneficial pumping scheme depends on the actual system configuration. In any case, the bi-directional pump scheme may result in higher gain saturation and it also allows using longer fibers which, as has been seen in section 6.1.3, tends to lead to higher TMI thresholds. Thus, when compared to the counter-pump scheme, some experiments have shown an increase of the TMI threshold by ~50 % [139] when using the bi-directional pump configuration (which allowed benefiting from a longer effective amplification length). This pump scheme is of interest especially for CW fiber lasers where the accumulated non-linearity of the system is relatively low. On the contrary, the bi-directional pump configuration is normally less attractive for ultrafast fiber laser systems due to the associated penalty in non-linearity. Therefore, for these systems the counter-propagating pump configuration is still usually preferred.

In a different context, bi-directional pumping is the configuration of choice when using single mode fibers [17,78,80] since it leads to the lowest peak heat-loads in the fiber and, therefore, it delays the appearance of guided HOMs in these waveguides (and, with it, the onset of TMI). Anyway, it should be stressed that this strategy, which pursues preventing the waveguides from becoming multimode, can only be exploited when using fibers that operate in the single-mode regime; for few mode fibers it is ineffective.

As soon as PD is present in the fiber, the counter-propagating pump configuration seems to be the most favorable one [74,84], mainly because the high-power signal propagates over a relatively short length of the fiber (just at the very end), which limits the PD absorption of the signal radiation and, therefore, the extra heat-load generated by this process. In a different context, some theoretical studies point out that using distributed pump schemes [5] might lead to higher TMI thresholds [135] but no experimental evidence has been presented yet.

On a different account, the pump wavelength decisively influences the performance of a fiber laser system. This parameter determines the fiber length, the fiber geometry/doping concentration and the amount of heat-load induced by the QD and the PD-losses. Since all these parameters, in turn, also influence the value of the TMI threshold, it is reasonable to assume that so must the pump wavelength. This assumption has been corroborated with some theoretical and experimental work [64,74,82]. Thus, for example, the expected evolution of the TMI threshold with the pump wavelength in the presence of PD, as done in [74], is shown in Fig. 24. These simulation are a thermal optimization of the fiber (i.e. they try to maximize the output power for a given heat-load) and the TMI threshold is defined as the average power at which a certain constant average heat-load (of 34 W/m in this case) is reached in the fiber. As mentioned above, even though this is a very simple approach and the actual gain/loss in terms of the TMI threshold might differ to a certain extent from the one calculated, the general behavior of this parameter with the pump wavelength is expected to follow the trend shown in

Fig. 24. The simulations of [74] were done for a fiber with a 63 µm core diameter (~58 µm mode field diameter -MFD- in the fundamental mode), ~50 µm doped diameter, an AFF = 0.5 and a bulk Yb-ion concentration (i.e. the ion concentration of each one of the doped islands in Fig. 19) of $7 \cdot 10^{25}$ ions/m$^3$. The fiber is seeded with 30 W at 1030 nm. Two different cases are shown in Fig. 24: a) for a fiber with a fixed length (of 1.2 m) in which the pump core diameter is adapted for each pump wavelength to keep the small-signal pump absorption constant; b) for a fiber with a fixed pump core diameter (of ~200 µm) in which its length is varied to achieve a constant small-signal pump absorption for different pump wavelengths. As can be seen in Fig. 24a, there seems to be a maximum of the TMI threshold at the pump wavelength of 976 nm. According to these simulations, this is due to the fact that the lower absorption cross-section at the other pump wavelengths implies that smaller pump core diameters have to be used to increase the overlap of the pump with the doped region. This, in turn, results in a higher PD-induced heat-load and reduces the TMI threshold.

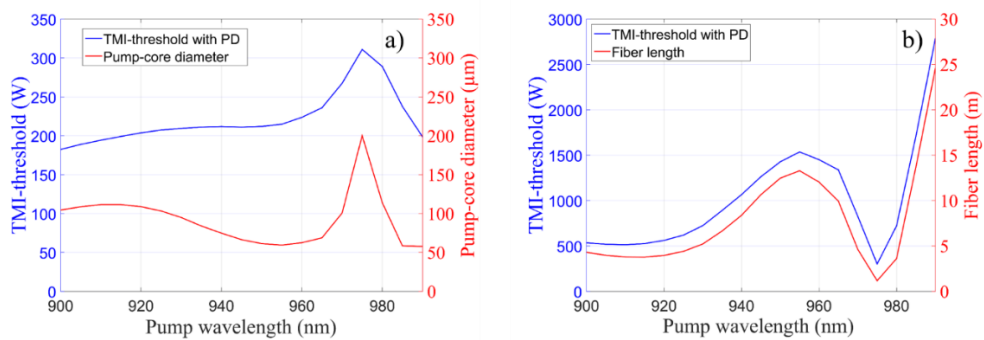

Fig. 24. Dependence of the TMI threshold on the pump wavelength (blue lines): a) for a 1.2m long fiber which pump core diameter (red line) is changed to achieve a constant small-signal pump absorption, and b) for a fiber with constant transverse dimensions which length (red line) is changed to achieve a constant small-signal pump absorption. Reprinted from Opt. Express 24, 7879 (2016) [74].

It has to be said, though, that there is currently no experimental evidence to corroborate or refute the simulation results presented in Fig. 24a, due to the difficulty in achieving comparability between fibers with different geometries. Nonetheless, the results presented in Fig. 24a seem to contradict the observations published in [64,140,141], where the authors report about higher TMI thresholds using pump wavelengths that are shorter or longer than 976 nm. However, even though in those experiments the physical fiber length was left constant (as done in Fig. 24a) the fiber length was deliberately chosen to efficiently absorb the pump at wavelengths other than 976 nm. In other words, the fibers were too long for the pump at 976 nm. Therefore, in the experiments, the effective length of the fiber (defined as the length of the fiber in which ~95 % of the pump is absorbed or in other words, where significant signal amplification takes place) changed with the pump wavelength. Plainly speaking the fiber was effectively longer for pump wavelengths different from 976 nm. Thus, this situation has also been simulated in [74] and the results are presented in Fig. 24b. In this set of simulations the transverse fiber dimensions are kept constant, but in order to ensure the comparability of the results, the fiber length (red line in Fig. 24b) is changed to obtain the same small-signal pump absorption at all pump wavelengths. This reveals a dependence of the TMI threshold on the pump wavelength that shows a minimum for a pump wavelength of 976 nm (as reported in the experiments). The reason is that the shortest fiber length corresponds to this pump wavelength (because the pump absorption cross-section reaches its peak value at 976 nm). In fact, the evolution of the TMI threshold (blue line) presented in Fig. 24b closely follows that of the fiber length (red line). This suggests that the dependence of the TMI threshold on the pump wavelength that has been reported in the experiments (specifically, it becoming higher for pump wavelengths different than 976 nm) is mostly due to the change of the effective length of the fiber (see section 6.1.3) created by changing the experimental conditions. Thus, as already discussed for the case of bi-directional pumping, changing the length is a viable mitigation strategy of TMI only for fiber laser systems not limited by non-linear effects. This means that, usually, this approach is not very attractive for ultrafast laser systems.

Another attractive approach to mitigate TMI can be that of tandem pumping [142,143]. Hereby a fiber laser (or several of them) is used to pump the Yb-doped fiber at wavelengths > 1000 nm (usually ~1018 nm [6]). It has been shown that this strategy can reduce the amount of photodarkening (due to the lower inversion levels achieved) in the fiber [144] as well as the quantum defect of the laser process (due to the long pump wavelengths). Thus, several theoretical studies predict that the use of tandem pumping should lead to higher TMI thresholds due to the lower heat-load generation in the fibers [54,134]. The main problem with tandem pumping is the relatively low absorption cross-section for the pump wavelength, which implies that longer fiber lengths have to be used, which is, again, only possible in systems not limited by non-linear effects. Besides, tandem pumping leads to low inversion levels in the fiber, which makes this strategy significantly less attractive for pulsed fiber laser systems (in particular for those trying to scale the pulse energy extracted from the fiber).

*6.2 Active mitigation strategies*

TMI is an effect in which a dynamic output is generated from a temporally stable input. Due to symmetry considerations it can be argued that, in such a system, a dynamic input could, in principle, lead to a temporally stable output. This is the basic idea behind active mitigation strategies of TMI. Hereby there is always some parameter (e.g. the coupling in the fiber, the pump power, or the signal power) that is modulated in a way that it achieves a stabilization of the output beam above the TMI threshold.

Even though passive mitigation strategies are generally preferred because they do not increase the complexity of the system, active strategies have already shown a greater scaling potential for the TMI threshold. For example, in the relatively short time since some of these methods have been presented, they have demonstrated the ability to increase the TMI threshold by factors > 2, something that is usually very hard to achieve with a passive mitigation strategy. Furthermore, since active mitigation strategies are usually perfectly compatible with passive ones, they offer a route to further increase the output average power of fiber laser systems once all the passive routes have been exhausted.

### 6.2.1 Dynamic mode excitation with an acousto-optic deflector

The first active mitigation strategy demonstrated stabilized the fluctuating beam above the TMI threshold by dynamically changing the mode excitation using an acousto-optic deflector (AOD) and a control-loop [145].

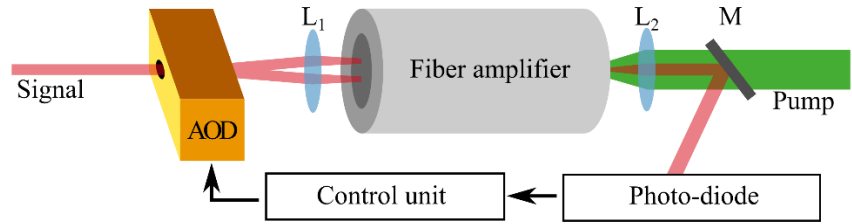

Fig. 25. Experimental setup of a fiber amplifier system with an acousto-optic deflector (AOD) used to stabilize the output beam above the TMI threshold. Adapted from Opt. Express 21, 17285 (2013) [145].

An AOD is an active optical element that deflects an input signal beam proportionally to an electric signal, i.e. the output propagation angle of the beam is changed. Such changes of the propagation angle of the beam can be translated into transversal position shifts with the help of a lens. Therefore, the coupling conditions in an active optical fiber can be dynamically changed with an AOD, in a controllable and reproducible manner. Then, by placing such a system into a feedback control loop, it is possible to devise a mechanism to stabilize the output beam. The experimental setup is schematically illustrated in Fig. 25, where it can be seen that the stability of the output beam is measured by a PtD (as explained in section 2) and then fed back to a micro-controller that, using this information, takes the decision on the next signal that should be applied to the AOD to stabilize the output beam.

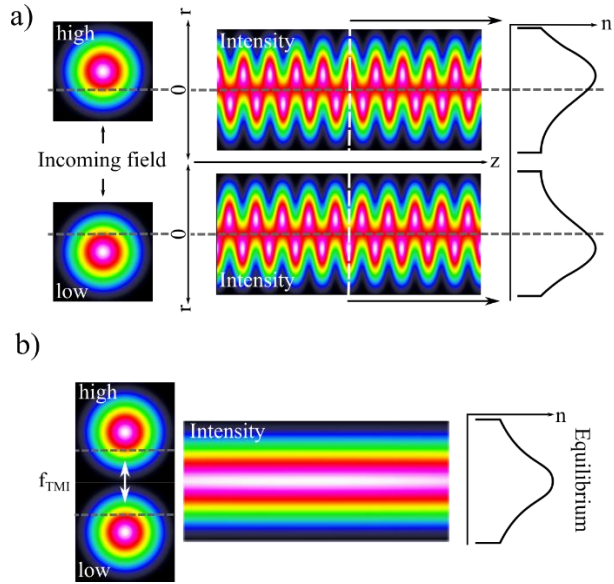

Fig. 26. Schematic representation of the operation principle of the TMI-stabilization system based on an acousto-optic-deflector. a) When the input beam is deflected from the upper to the lower part of the core (with respect to the optical axis of the fiber) in a quasi-static manner, it generates a new modal interference intensity pattern and thermally-induced index grating that is π-phase shifted with respect to the old one (i.e. that corresponding to the beam being deflected to the upper position). b) If the deflection between the upper and lower positions of the core is done with the right frequency ($f_{TMI}$), then a balance between both thermally-induced gratings is achieved resulting in an index change that is radially symmetric and which, therefore, will only cause a weak coupling between the transverse modes. Adapted from Opt. Express 21, 17285 (2013) [145].

The operating principle of this method is schematically described in Fig. 26. For the sake of simplicity it will be assumed that the AOD can only switch the beam between two positions situated symmetrically with respect to the fiber axis. Additionally it will be assumed that only two fiber modes are excited by the incoming beam: a radially symmetric FM and a radially anti-symmetric HOM. Thus, when the beam is deflected to the upper position (Fig. 26a, upper part), then a certain modal interference intensity pattern is created. On the other hand, when the beam is switched to the lower part of the core (Fig. 26a, lower part) then the exact same MIP is generated only with a π-phase shift (i.e. flipped around the fiber axis). In other words, the new intensity maxima fall at the positions where the former intensity minima were located and vice-versa. Such a π-phase shift of the MIP occurs always when the center of gravity of the beam crosses the fiber axis and it is independent of the exact position of the beam. Thus, by deflecting the incoming beam it is possible to induce a π-phase shift of the MIP and, therefore, of the thermally-induced index grating. If this deflection, or in other words, if the alternation between the two positions of the beam happens too slowly, then the RIG has enough time to adapt itself to the new MIP, as schematically shown in Fig. 26a, and ends up allowing for the energy transfer between the modes (provided there is a mechanism inducing a phase shift between MIP and RIG, as discussed in section 5). On the contrary, if the deflection is too fast, then the thermally-induced index grating will adapt itself to the average of the MIP, which will most likely be transversally inhomogeneous (if the two beam positions in the cores are not perfectly symmetric) and, therefore, will allow for the energy transfer between the FM and the HOM (again if there is a phase shift). Only in the case in which the deflection frequency is right, it is possible to get an equilibrium state in which the RIG starts to change but doesn't have enough time to adapt itself fully to the new MIP. As a result, the RIG is washed out transversally and becomes significantly more radially symmetric, which strongly reduces the coupling efficiency between the radially symmetric FM and the radially anti-symmetric HOM (Fig. 26b).

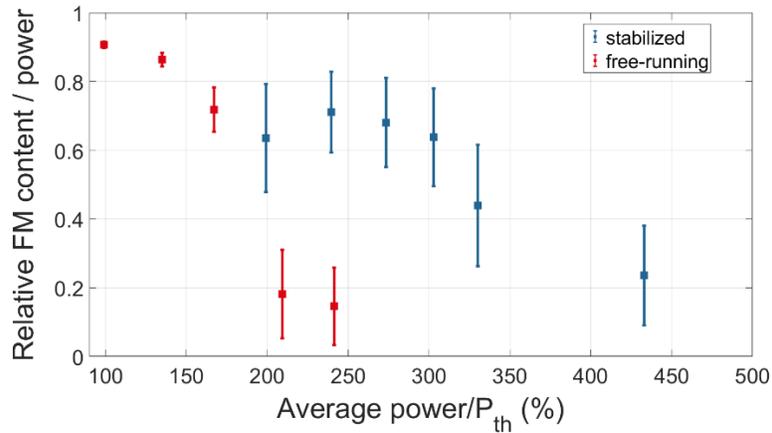

Fig. 27. Experimental results obtained for the active stabilization of TMI with dynamic mode excitation by using an acousto-optic deflector (AOD). The plot shows the evolution of the relative modal power in the FM as a function of the output power (normalized to the TMI threshold power $P_{th}$) with (blue points) and without (red points) AOD. Adapted from Opt. Express 21, 17285 (2013) [145].

The results of the beam stabilization are summarized in Fig. 27. This plot shows the evolution of the relative modal power contained in the FM (determined from the modal reconstruction of a high-speed video) as a function of the output average power (normalized to the TMI threshold power $P_{th}$) with (blue points) and without (red points) the stabilization system. For powers above 250 % of the TMI threshold it was not possible to measure the FM content of the free-running system anymore. That is why no red points are plotted beyond this point. Fig. 27 clearly shows that the FM-content can be significantly increased when using the stabilization system, especially in the power range between 200 % and 300 % of the TMI threshold of the free-running system. Beyond that power, as more and more HOMs take part in the beam fluctuations, the relative content of the FM steadily drops.

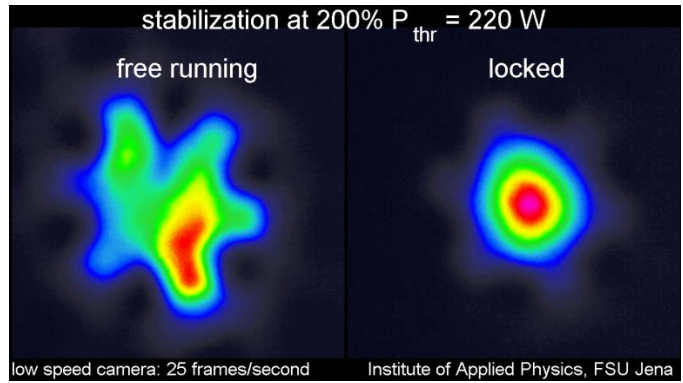

Fig. 28. Frames from a video showing the comparison between the free running (left) and the stabilized beam profile (right) at an average power that is two times higher than the TMI threshold power $P_{thr}$ (220 W). Reprinted from Opt. Express 21, 17285 (2013) [145].

The result of the stabilization can be qualitatively evaluated in Fig. 28, where two frames from a video showing the impact of this stabilization technique (at a power that is two times above the TMI threshold) are shown. Thus, the beam of the free-running system (i.e. with the AOD switched off) can be seen on the left-hand side, while the stabilized beam (i.e. with the AOD switched on) can be seen on the right-hand side. This picture clearly shows the significant improvement in beam quality achieved by the AOD.

This experiment shows that an external stabilization can be effectively used to mitigate the effects of TMI. However, this strategy leads to an increase in the overall complexity of the system. Even though the results presented herein are good enough to illustrate the effectiveness of the technique, they are only a proof-of-principle experiment that is far from being optimized. Therefore, it can be expected that an even better stabilization can be achieved with a further improvement of the control loop.

### 6.2.2 Dynamic mode excitation with a photonic lantern

A refinement of the TMI mitigation strategy presented in the previous subsection has been recently published [146]. Hereby the AOD is replaced by a photonic lantern [147–149], which offers more degrees of freedom for the excitation of the active fiber. In fact, photonic lanterns allow individually controlling the amplitude, phase and polarization of different fiber modes. This makes it possible to synthesize a rich variety of beams to excite an active fiber using these optical elements.

The extra flexibility offered by the photonic lantern with respect to the AOD, was put to good use in [146]. This work describes a kW-level fiber MOPA system in which the input end of the main active fiber (25 µm core diameter, 400 µm cladding diameter, 12 m long) was spliced to a photonic lantern. The photonic lantern had three input ports, each one incorporating amplitude, phase and polarization modulators. By acting upon these modulators it was possible to control the beam injected in the active fiber. Thus, the authors of the work used a stochastic parallel gradient descent (SPGD) algorithm [150] to dither the signal applied to the modulators and, based on the signal captured by a pinhole photodetector that sampled the output beam, maximize the on-axis response of the output light.

The free-running system did show some beam fluctuations starting from a power of 800 W, but it is still unclear whether these were due to TMI or not, since the fluctuations were significantly slower than what could be expected from TMI. Regardless of the origin of the beam fluctuations, to which the stabilization method is agnostic, the system could be successfully stabilized up to a pump-limited output power of 1.27 kW. This results has demonstrated the feasibility of active beam stabilization in an all-fiber configuration.

### 6.2.3 Pump modulation

Another way to stabilize the beam above the TMI threshold is by modulating the pump power [52]. In fact, this approach is complementary to the one that uses the AOD (see subsection 6.2.1). The advantage of this technique is its simplicity and the fact that in can be readily implemented in an already deployed fiber systems. As a matter of fact, the only thing that is necessary to implement this approach is a pump laser driver that accepts an external modulation signal, as seen in Fig. 29. It is worth highlighting that no control feedback loop had to be implemented in this setup, which clearly contrasts with the one employing the AOD (see Fig. 25). This not only simplifies the setup, but also highlights the robustness of this beam stabilization approach.

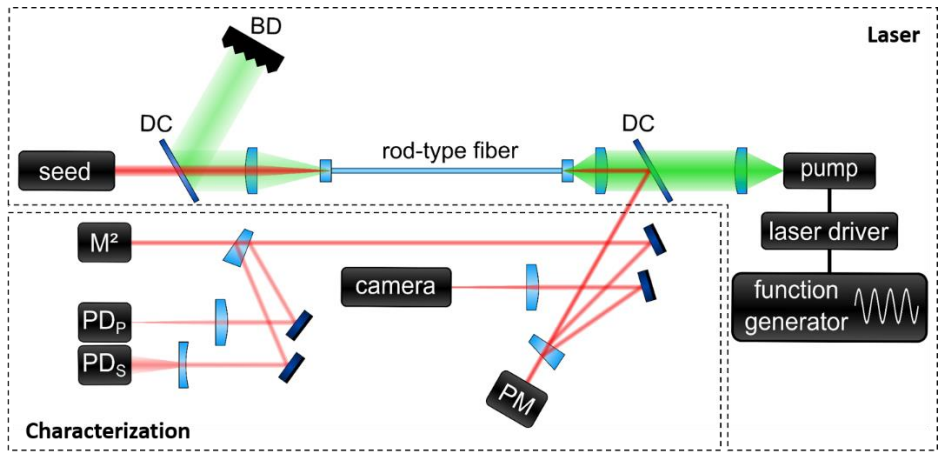

Fig. 29. Experimental setup used to mitigate TMI with pump modulation. This setup can be divided in two distinct parts: the laser itself and the characterization part. Here the different components are: DC - dichroic mirror, BD - beam dump, PM - power meter, PD – photodiode, $M^2$ -beam quality characterization. Adapted from Opt. Express 26, 10691 (2018) [52].

The main idea behind the pump modulation approach is similar to that of the AOD: to wash out the RIG. However, in this case the RIG will be washed out in the longitudinal direction (instead of in the transverse direction as done with the AOM). This principle is illustrated in Fig. 30. As can be seen, when the pump is increased (decreased), the MIP will be compressed (stretched), due to the higher (lower) heat-load generated in the active fiber. This heat-load change alters the transverse refractive index profile of the fiber via the thermo-optic effect. This, in turn, changes the propagation constants of the fiber modes, which eventually modifies the MIP. Thus, if the pump is modulated, the MIP will be constantly stretched and compressed, describing a movement akin to that of an accordion. When this is done with an appropriate frequency, it should lead to a washing out of the RIG in the longitudinal direction as the temperature profile tries to follow the shift of the MIP.

At this point is it worth noting that, with an actual modulation of the pump, the stretching/compression of the MIP will be more subtle than that shown in Fig. 30. The reason is that the simulations presented in the figure were carried out for the CW case, i.e. the temperature profile has infinite time to change. However, in reality, during the pump modulation the temperature profile has only a limited time to evolve from one state to the other. Thus, the higher the modulation frequency, the less movement will be induced in the MIP. This already suggests that the pump modulation scheme should have a certain dependence on the modulation frequency. In fact, if the modulation frequency is too high, the MIP will barely move and, therefore, no washing out of the RIG will occur. Alternatively, if the modulation frequency is too low (i.e. if the movement of the MIP happens in time scales significantly larger than the thermalisation time of the fiber core), there is enough time for the RIG to change and to follow the movement of the MIP and, thus, it will not be washed out either. Therefore, there is a "sweet spot" for the modulation frequency at which the RIG has enough time to start adapting to the shift of the MIP but it cannot fully follow its movement. Thus, the RIG finds itself in a constant transition state (continuously moving up- and downstream the fiber without ever reaching its steady-state profile for the maximum or the minimum pump) that results in its weakening, i.e. it is washed out. Such weakening of the RIG leads to a weaker modal energy transfer and, therefore, to the stabilization of the beam even beyond the TMI threshold.

The "sweet spot" modulation frequency depends on the specific fiber (and, in particular, on its dimensions) used in the experiments, but it has been reported in [52] that this approach seems to have a large tolerance to the modulation frequency, i.e. that there is not a single frequency that works but a frequency range with a bandwidth of several hundred Hz.

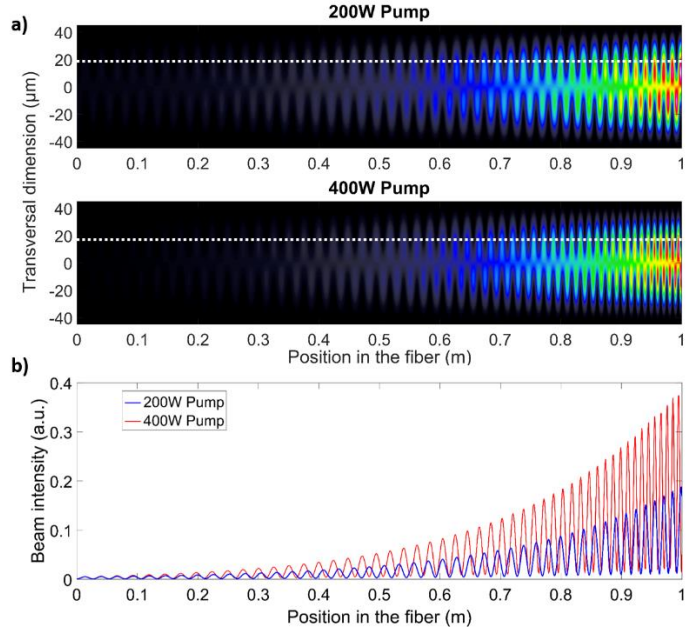

Fig. 30. a) Modal interference pattern along a 1 m long fiber for 200 W (upper graph) and for 400 W (lower graph). b) MIP along a line parallel to the fiber axis and shifted 20 µm from it (white dotted lines in a)). A change of the pump power results in a stretching or compression of the period of the MIP. Reprinted from Opt. Express 26, 10691 (2018) [52].

According to section 5, a modulation of the pump also leads to a phase shift between the MIP and the RIG and, for high enough average powers, to modal energy transfer. Therefore, it can be expected that this also happens to a certain extent in this mitigation strategy. At this point it is worth stressing, however, that the results presented in section 5 and, in particular, in Fig. 13 are not representative of the amount of modal energy transfer obtained with the modulation of the pump in this mitigation strategy. The experiment in Fig. 13 was conceived to purposefully maximize the energy exchange between the modes by using a modulation frequency that is significantly lower than the one required to wash the RIG out. Actually, as discussed in [52] and shown in [43], for the modulation frequencies that result in a successful washing out of the RIG, such an energy transfer is weak and is restricted to a brief portion of the modulation period. In fact, according to [43], the increase in the HOM content induced by this energy transfer should be lower than 5 % (integrated over one modulation period), if the modulation frequency is properly chosen. In other words, even though some amount of energy transfer is unavoidable when using this active mitigation strategy, it will have a negligible impact for most applications.

This TMI mitigation strategy has been experimentally tested in a fiber laser system with a 1.1 m long active rod-type fiber (LPF) with an active core of ~65 µm. This fiber was seeded by a 5 W signal at 1030 nm and counter-pumped by a 976 nm laser diode which, connected to a driver, could be modulated with frequencies up to ~1 kHz. The TMI threshold of the free-running system was reported to be 266 W (measured using the method described in section 2). The system was then operated at 407 W (factor 1.5 above the TMI threshold) and the output beam was characterized with the setup shown in Fig. 29. This means that, among other things, the output beam was imaged and recorded with a CCD camera (30 fps). Fig. 31a shows a frame captured by this camera when the pump modulation is switched off (i.e. for the free-running system). There is a clear presence of HOM in this beam profile. When a pump modulation of 720 Hz with a modulation depth of ±77 % was applied, TMI could be mitigated, which resulted in the stable beam shown in Fig. 31b. Moreover, it has been reported that the $M^2$ of the beam improved from > 1.6 for the free-running system to ~1.1 with the stabilized beam. Using this strategy it was possible to stabilize the beam up to an output power of ~560 W, which corresponds to a power more than 2 times above the TMI threshold.

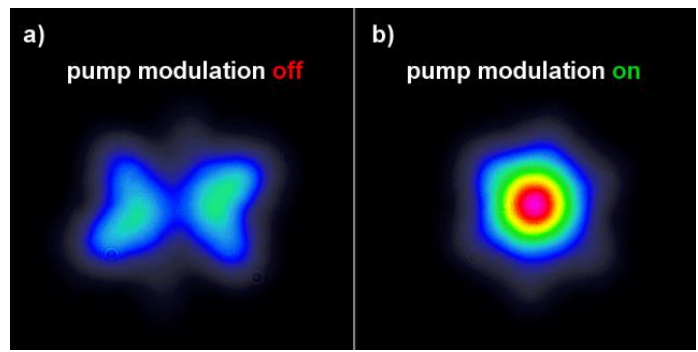

Fig. 31. Beam profiles at 407 W of output average power (a factor of 1.5 above the TMI threshold) emitted by a) the free-running system and b) the system with the pump modulation switched on. Reprinted from Opt. Express 26, 10691 (2018) [52].

The main drawback of this approach to mitigate TMI is that, due to the changes in the pump power, the output power is also (strongly) modulated. Even though this fact could be tolerated by some applications (especially those with interaction times longer than the modulation period), it is certainly a significant problem for many others. This drawback reduces the usability and attractiveness of this approach. Even though it has been argued that, quite likely, the modulation depth required to mitigate TMI in long fibers will be much lower than the one used in [52] and this will certainly palliate the problem in those systems, it will not complete resolve it. In this context, a solution has been proposed in the form of a system with a bi-directional pumping scheme in which each pump source is modulated with a temporal π-phase shift to one another. This way the output power will be kept constant but the MIP will still be shifted up- and downstream the fiber (since the position of the highest heat-load will shift). However, no experimental demonstration of this refinement has been reported yet.

6.2.4 Phase-shift manipulation with seed bursts

As already mentioned in section 5, a new family of TMI mitigation strategies based on the control of the phase shift between the MIP and the RIG has been recently proposed. The first representative of this family of techniques is one that exploits burst operation to achieve a positive phase shift (as defined in Fig. 11) in the fiber amplifier [97].

It has been seen in section 5 that when the output power increases, the MIP is compressed because the temperature rises, which leads to a positive phase shift and, with it, to a positive energy transfer. This effect resulted in a cleaning of the output beam in the rising edges of the pump modulation cycle even above the TMI threshold, as described in [43]. This is the idea exploited with the seed burst approach: to generate transients of emission in which the temperature rises, which leads to a positive energy transfer, i.e. to a beam cleaning. However, since the temperature in the fiber cannot increase perpetually, the emission is switched off after a certain time, allowing the fiber to cool down before the next emission burst comes. The length of the burst should be such that it does not allow the phase shift to grow beyond π.

In order to test the feasibility of this approach, some simulation have been performed using the model described in [52]. The simulations have been carried out for a 1 m long fiber with 80 µm core diameter (V-parameter 7), 228 µm pump core diameter, doped with $3.5 \cdot 10^{25}$ Yb-ions/m$^3$ and seeded by 30 W of signal power (10 % of which is in the HOM) at 1030 nm. This fiber is pumped at 976 nm and the pump is progressively increased over the first milliseconds of the simulation until it reaches its final steady-state value. Two cases have been simulated with this fiber: a free-running system emitting 300 W of average power and a burst system (400 Hz burst frequency and 50 % duty cycle) also emitting 300 W of average power.

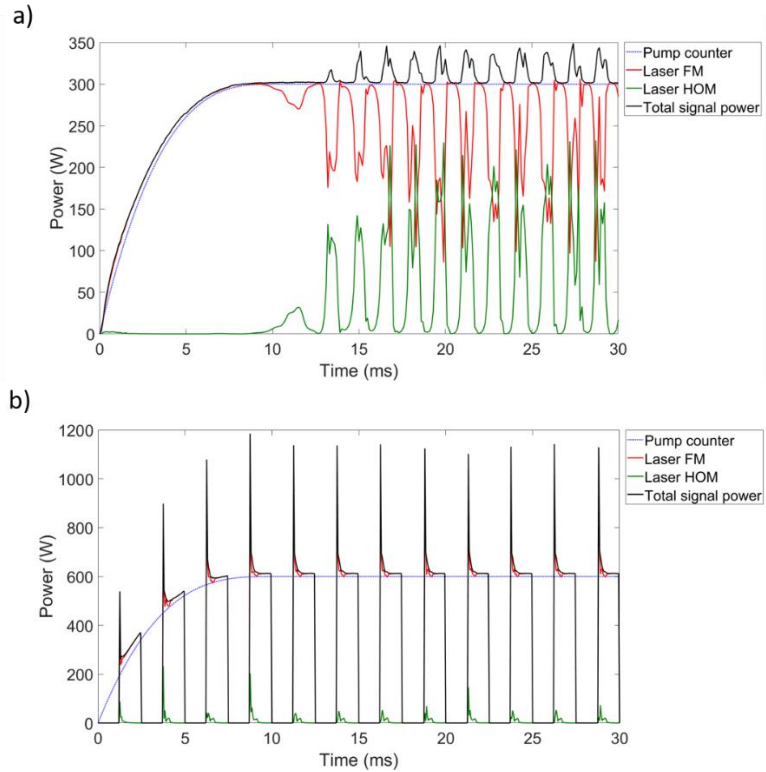

Fig. 32. Temporal evolution of the pump (blue), FM power (red), HOM power (green) and total signal power (black): a) for the free-running system with 300 W output average power and b) for the system using seed-power bursts with 50 % duty cycle and emitting 300 W output average power.

The results for the free-running system are presented in Fig. 32a. There it can be seen that, as the pump power increases (dotted blue line), the total output power also increases (black line) until the steady state is reached. It is interesting to note that the emitted power is almost exclusively contained in the FM (red solid line) during the ramp up of the pump power. However, after a few ms of the system operating at its steady-state value, there is an onset of strong modal fluctuations in which the energy flows back and forth between the FM and the HOM (green solid line). This reveals that the system is operating above the TMI threshold. In fact, according to the simulations, the threshold is around 200 W in this case, so the system operates roughly 1.5 times above it. It is worth noting that the periodic changes of the total output signal power are a consequence of the modal fluctuations and they arise due to the different overlaps of the modes with the active region.

The result of the simulation looks dramatically different when using the seed bursts, as seen in Fig. 32b. Please note that only the envelopes of the burst can be observed in the figure (black lines). This means that the system may have a significantly faster pulse train underneath these burst envelopes (or it can be operated in quasi-CW mode using these long pulses, instead). In any case, it can be observed that the energy is mainly contained in the FM during the burst duration. There is, nonetheless, a small percentage of the energy contained in the HOM for a brief lapse of time at the very front of each seed burst. The reason for this is that a certain amount of time is required to generate the positive phase shift in the system and, therefore, to start the beam-cleaning process which, in turn, transfers the energy from the HOM into the FM, as can be seen in Fig. 32b. This results in a stable emission with nearly diffraction-limited beam quality at an average power 1.5 times above the TMI threshold. In fact, simulations predict that this mitigation strategy should work for powers up to 2 times above the threshold.

The simulations presented in Fig. 32 seem to confirm the effectiveness of this strategy for the mitigation of TMI. However, there are a couple of drawbacks associated with this technique that should be mentioned. First of all, the parameters of the burst operation cannot be freely chosen. This means that there is a certain combination of burst duration and duty cycle that offers the best mitigation performance [97]. This lack of flexibility only affects the burst envelopes, the parameters of the pulse train underneath them can be changed at will. Such a relatively low flexibility may be problematic for some applications. Another drawback is the fact that the front of each signal burst has a very high power spike, which might damage the system and/or be detrimental for the application. However, in principle such a spike can be flattened out by shaping the input seed bursts. Finally, another drawback of the technique is that, in the example discussed above, the system was pumped with 600 W but the average power emission was still just 300 W, due to the 50 % duty cycle of the seed bursts. This means that there is a significant drop of net amplification efficiency when using this technique. However, this can be partially relativized when realizing that the intra-burst output power is also ~600 W. This means that the pulses being emitted (in a stable manner) have a peak power that is 3 times above the TMI threshold. Thus, using this approach, the gain in peak power/pulse energy is even higher than the gain in average power, which is a very attractive feature for some applications. Additionally, this loss of amplification efficiency could be further mitigated by pulsing the pump with the same period as the seed burst.

It is worth mentioning that the mitigation strategy described in this subsection is only the very first proposal of a technique aiming at exploiting the physics (particularly, the impact of the phase shift) behind TMI to improve the performance of a fiber laser systems. Thus, it is expected that new, more refined representatives of this very promising family of mitigation strategies will be presented in the future. These can potentially revolutionize fiber laser technology by ensuring that high-power fiber laser systems emit a beam in the FM (almost) independently of the excitation and fiber design.

## 7. Conclusion

The effect of transverse mode instability (TMI) is the first representative of thermally-induced non-linear effects in optical fibers. This phenomenon was first reported in 2010 and it has become the strongest limitation for the scaling of the average power of fiber laser systems in a very short time. Even worse, TMI undermine the excellent reputation of fiber laser systems, which is grounded on their ability to emit a power-independent beam quality. As a consequence, there is a spreading fear that TMI may represent the ultimate limit for the average power of fiber laser systems. Therefore, any research result related to this effect is always received with a lot of interest and expectation.

In this work we have summarized the research activities on the topic of TMI that have been carried out worldwide in the last decade. These encompass all the different aspects related to the phenomenon: from the first experimental observations, to the theoretical description of the effect or the presentation and the successful demonstration of mitigation strategies.

In this work the first observations of TMI and the first systematic experimental characterization of the effect have been described. Thus, it could be shown that there are three main operation states in a fiber laser system: stable, transition and chaotic states. Additionally, it could be experimentally confirmed that TMI leads to an energy exchange between two or more transverse modes of a fiber. Furthermore, it has also been shown that the fluctuation frequency of the beam instabilities is compatible with the thermal origin of the effect.

Using these observations as a basis, the most widely accepted physical description of the process has been presented: two transverse modes in a fiber create an interference pattern (MIP) that, through the thermo-optic effect, gives rise to a refractive index grating (RIG). This grating, in turn, has the right characteristics to potentially transfer energy between the interfering modes. However, for this energy transfer to take place, it is necessary that a longitudinal phase shift between the MIP and the RIG evolves. Moreover, recent experiments have demonstrated that the sign of this phase shift determines the direction of the energy transfer, i.e. whether the energy flows from the FM to the HOMs or the other way around. Furthermore, the newest experimental evidence indicates that pump-power fluctuations can lead to such phase shifts and, therefore, pump-intensity noise can be responsible for triggering TMI in high-power fiber laser systems. These interactions have been described by different models used to simulate TMI, which have been briefly described in this work.

It has also been shown that photodarkening (PD) has a strong impact on the TMI threshold. PD is a well-known phenomenon which happens in Yb-doped fibers and which has been investigated for more than a decade now. Prior to the appearance of TMI, PD was considered to be nearly resolved for most practical applications or, at least, under control. However, the research on TMI has shown that even a low level of PD can still significantly influence the TMI threshold through the non-negligible amount of extra heat generated in the fiber. This finding leads to the need for completely rethinking the approach used to optimize PD: the active medium should not only be optimized for low optical losses but also for a low heat-load. As a consequence of this change in paradigm, some guidelines to optimize the material of a fiber have been recently proposed. The implementation of some of these guidelines is already leading to the demonstration of a significant increase of the average power emitted by diffraction-limited fiber laser systems.

Finally, several mitigation strategies for TMI have been presented and discussed. These comprise both passive and active techniques. The first type of strategies involves changes in the system (such as fiber design, pump/signal wavelength, pump direction, gain saturation, etc.) whereas the second type requires some dynamic control of the system (e.g. of the coupling conditions, of the pump power, etc.). Traditionally most mitigation strategies for TMI act upon the strength of the RIG trying to weaken it by various means. However, as has been reviewed in this work, a new family of techniques has been recently proposed, which try to control the direction of the energy transfer by acting upon the phase shift between the MIP and the RIG. This family of mitigation strategies tries to exploit the physics driving TMI to improve the performance of high-power fiber laser systems.

Even though TMI is presently still the strongest limitation for the further scaling of the average power of fiber laser systems, the rapid development of the understanding of this effect gathered worldwide is very encouraging since it paves the way to effectively mitigating or even exploiting TMI in the near future. At the light of the newest experimental results it is reasonable to expect that the average power of diffraction-limited fiber laser systems starts rising again. Such prospects let the future of fiber laser technology look bright again.

## Acknowledgements

This work has been funded by the Deutsche Forschungsgemeinschaft (DFG, German Research Foundation) - 416342637; 416342891; GRK 2101 (259607349). Fraunhofer Gesellschaft – Fraunhofer Cluster of Excellence "Advanced Photon Sources".

**Biographies**

**Dr. Cesar Jauregui** was born in Santander, Spain, on June 18, 1975. He received both his Telecommunication Technical Engineering degree and his Telecommunication Engineering degree at the University of Cantabria. In 2003, he got his Ph.D. degree at that same University. In 2005 he began a two-year post-doc stay at the Optoelectronics Research centre, where he investigated the phenomenon of slow-light in optical fibers. Since 2007 he is working at the Institute of Applied Physics in Jena. His primary research concerns are high-power fiber lasers, non-linear effects and mode instabilities in optical fibers.

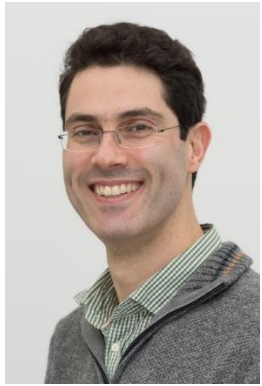

**Christoph Stihler** was born in 1990 in Leipzig, Germany. He studied "Laser- and Optotechnologies" at the University of Applied Sciences in Jena and received his Master's degree in 2015. Since 2015 he has been working as a PhD student in the fiber lasers group of the Institute of Applied Physics at the Friedrich-Schiller-University Jena in the field of transverse mode instability in high-power fiber lasers.

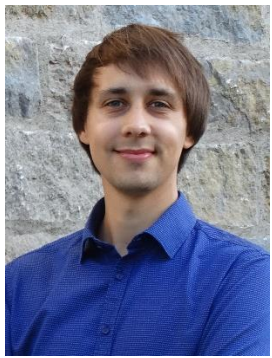

**Jens Limpert** was born in Jena, Germany, on 11th of December in 1975. He received his M.S in 1999 and Ph.D. in Physics from the Friedrich Schiller University of Jena in 2003. His research interests include high power fiber lasers in the pulsed and continuous-wave regime, in the near-infrared and visible spectral range. He has published over 50 conference and journal papers in the field of fiber lasers. After a one-year postdoc position at the University of Bordeaux, France, where he extended his research interests to high intensity lasers and nonlinear optics, he returned to Jena and is currently leading the Laser Development Group (including fiber- and waveguide lasers) at the Institute of Applied Physics.

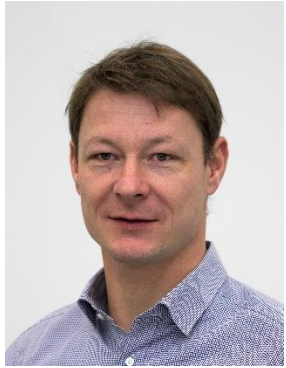